\documentclass[11pt]{article}

\newcommand{\blind}{1}

\usepackage{dsfont}
\usepackage{bbm}
\usepackage{color}
\usepackage[ngerman,english]{babel}
\usepackage{graphics}
\usepackage{graphicx}
\usepackage{epsfig}
\usepackage{lscape}
\usepackage{amsmath,amsfonts,amssymb}
\usepackage{eurosym}
\usepackage{natbib}
\usepackage{url}
\usepackage{hyperref}
\usepackage{setspace}
\usepackage{caption}
\usepackage{subcaption}
\usepackage{arydshln}
\usepackage{rotating}
\usepackage{floatflt}
\usepackage{pdflscape}
\usepackage{fullpage}

\newtheorem{theorem}{\small\sc Theorem}
\newtheorem{proposition}{\small\sc Proposition}
\newtheorem{assumption}{\small\sc Assumption}
\newtheorem{lemma}{\small\sc Lemma}
\newtheorem{definition}{\small\sc Definition}
\newcommand{\indep}{\perp\!\!\!\!\perp}
\newcommand{\E}{\mathbb{E}}
\newcommand{\bc}{\textbf{c}}
\newcommand{\Kr}{\widetilde{\mathbf{K}}^{(r)}}

\begin{document}
\def\spacingset#1{\renewcommand{\baselinestretch}%
{#1}\small\normalsize} \spacingset{1}


\if1\blind
{
  \title{\bf \Large Kernel Balancing: A flexible non-parametric weighting procedure for estimating causal effects}
  \author{Chad Hazlett\\
    \normalsize Departments of Statistics \& Political Science, UCLA}
  \bigskip
  
  \date{This version: 25 April 2016. \\ Based on PhD thesis, May 2014, MIT.}
  \maketitle
} \fi
 
\if0\blind
{
  \bigskip
  \bigskip
  \bigskip
  \begin{center}
    {\Large \bf Kernel Balancing: A flexible non-parametric weighting procedure for estimating causal effects}
\end{center}
  \medskip
} \fi

\bigskip
\begin{abstract}
In the absence of unobserved confounders, matching and weighting methods are widely used to estimate causal quantities including the Average Treatment Effect on the Treated (ATT). Unfortunately, these methods do not necessarily achieve their goal of making the multivariate distribution of covariates for the control group identical to that of the treated, leaving some (potentially multivariate) functions of the covariates with different means between the two groups. When these ``imbalanced'' functions influence the non-treatment potential outcome, the conditioning on observed covariates fails, and ATT estimates may be biased. Kernel balancing, introduced here, targets a weaker requirement for unbiased ATT estimation, specifically, that the expected non-treatment potential outcome for the treatment and control groups are equal. The conditional expectation of the non-treatment potential outcome is assumed to fall in the space of functions associated with a choice of kernel, implying a set of basis functions in which this regression surface is linear. Weights are then chosen on the control units such that the treated and control group have equal means on these basis functions. As a result, the expectation of the non-treatment potential outcome must also be equal for the treated and control groups after weighting, allowing unbiased ATT estimation by subsequent difference in means or an outcome model using these weights. Moreover, the weights produced are (1) precisely those that equalize a particular kernel-based approximation of the multivariate distribution of covariates for the treated and control, and (2) equivalent to a form of stabilized inverse propensity score weighting, though it does not require assuming any model of the treatment assignment mechanism. An \texttt{R} package, \texttt{KBAL}, is provided to implement this approach.

\end{abstract}

\noindent%
\vfill

\newpage

\section{Introduction}
Estimation of causal effects from observational data is a common goal of research endeavors across many disciplines, especially in the social sciences where many treatments of potential interest cannot feasibly be randomized. A widespread strategy for causal inference from such data involves first arguing that there are no \textit{unobserved} confounders, then adjusting the sample to make treated and control groups as similar as possible on \textit{observed} characteristics. Once this is done, a difference in mean outcomes or another outcome model can be run, and the remaining differences are assumed to be due to treatment rather than any effect of the covariates, as the covariates distribution has been made similar between the groups. Using such balancing procedures  prior to running outcome models is often preferable to only running an outcome model, particularly when the control and treatment groups have very different distributions (see e.g. \citealp{ho2007matching}), as illustrated in the applied example used here.

Kernel balancing is a sample-adjustment procedure of this type, though it's primary motivation targets a different goal than procedures such as propensity score methods, matching, and covariate-balancing weighting approaches. It proposes to use a high-dimensional choice of basis expansion on the original covariates, such that (a) the conditional expectation of the non-treatment potential outcome is assumed to be approximately linear in these bases, and (b) weights can be chosen that produce the same means for the treatment and control group on these bases, by using a kernel representation that makes such balancing tractable even for high or infinite dimensional expansions. Obtaining equal means on the these features/bases implies that the mean non-treatment potential outcome is also equal for the treated and control groups. Unbiasedness of a difference in means estimator for the average treatment effect on the treated (ATT) follows. 

Kernel balancing makes several contributions relative to existing matching and balancing approaches. First, while matching, weighting for covariate balancing, and propensity score methods can be understood as seeking to make the multivariate distribution of the covariates for the controls identical to that of the treated units, this is more than is required for unbiased estimation of the ATT. The goal of kernel balancing is, instead, to ensure that the non-treatment potential outcome has the same mean for the treated and control group under the most general conditions possible. One contribution of this paper is simply to emphasize that this simpler condition is all that needs to be met, and draw out its implication for procedures that seek covariate balancing weights. 

Second, in the matching and covariate-balancing literatures there is no clear answer to the important question of ``on what functions of the covariates should the investigator seek balance, i.e. equal means for the treated and control groups?'' Yet, different choices lead to different estimates, many estimators become infeasible for high-dimensional choices, and the failure to obtain balance on functions of the covariates can bias ATT estimates when that function influences the outcome (as illustrated in \ref{motivation} below). Leaving investigators with the choice of what functions to balance on poses significant challenges for the transparency and reliability of reported results as well. Kernel balancing ensures the treated and control groups will have the same means not only on the covariates, but on a wide range of flexible functions of the covariates, with little or no user intervention. While kernel balancing does not claim to be the only answer to the question of ``what functions to check balance on'', it provides one principled answer and the tools to implement it. 

Third, the use of kernels in choosing what functions to achieve balance on reveals an intimate relationship between the goals of achieving mean balance on these basis functions and the more traditional goal of achieving multivariate density equality. Specifically, the weights that achieve equal means on the basis functions associated with kernel $k$ and that are estimated by kernel balancing are precisely those that achieve equal \textit{estimated} joint densities of the covariates for the treated group and control group, where those estimates are formed using the same choice of kernel, $k$.  Thus, if a Gaussian kernel is used and we obtain equal means on the basis functions associated with this kernel via the procedure described below, then using those weights also implies that a Gaussian-based kernel density estimator of the covariate distributions will be equal for the treated and control groups. Following directly, I also show that these weights are equivalent to a non-parametric form of stabilized inverse propensity score weights, but without requiring a model for the propensity score. Thus, while focusing first on the minimum requirement for unbiased ATT estimation, the method also achieves an approximation to the multivariate density balancing goals for which matching, weighting, and propensity score have traditionally been employed.

In what follows, Section \ref{motivation} first provides an illustration of the risk of bias under existing methods and briefly previews the benefits of the proposed solution. Section \ref{proposal} details the idea behind kernel balancing. The discussion in Section \ref{discussion} places this approach in the context of existing matching, weighting, and propensity score methods, provides further properties of the method, and gives additional implementation details. Section \ref{empirical} tests the methods by using observational data to recover the experimentally determined effect of a well-known job training program on income (\citealp{lalonde1986evaluating,dehejia1999causal}). Further applications, proofs, and additional details are available in the Appendix.

\section{Motivation for the Method}\label{motivation}
I begin with a motivating example. While this is a simple simulation to highlight the practical challenges of existing methods and need for an alternative approach, the choice of variables and hypothetical relationships among them uses a real world substantive example so as to maintain clarity on its practical implications. 

Suppose we are interested in the question of whether peacekeeping missions deployed after civil wars are effective in lengthening the duration of peace ($peace\, years$) after the war's conclusion (e.g. \citealp{fortna2004does,doyle2000international}). However, within the set of civil war cases constituting our sample, the ``treatment'' -- peacekeeping missions ($peacekeeping$) -- is not randomly assigned. Rather, missions are more likely to be deployed in certain situations, which may differ systematically in their expected $peace\,years$ even in the absence of a peacekeeping mission. To deal with this, we collect four pre-treatment covariates that describe each case: the duration of the preceding war ($war\,duration$), the number of fatalities ($fatalities$), democracy level prior to the peacekeeping mission ($democracy$), and a measure of the number of factions or sides in the civil war ($factionalism$). We are interested in the ATT, which is the mean number of $peace\,years$ experienced by countries that received $peacekeeping$, minus the average number of $peace\,years$ for this group had they not received peacekeeping missions. 

Further, suppose there are no unobserved confounders, and that peacekeeping missions are deployed on the basis of a conflict's $intensity$, which equals $\frac{fatalities}{war\,duration}$. In particular, missions are more likely to be deployed where conflicts were higher in intensity. Suppose the outcome of interest, $peace\,years$, is also a function of $intensity$, with more intense conflicts leading to longer average $peace\,years$. This is reasonable if, for example, more intense wars indicate greater dominance by one side, leading to a lower likelihood of resurgence in each subsequent year. In this example, $peace\,years$ is only a function of $intensity$, and not of $peacekeeping$, implying a true treatment effect of zero. See Appendix \ref{motivationdgp} for the complete description of the data generating process. 

\begin{figure}[!hbt]
  \centering
  \caption{Imbalance on a function of the covariates}
  \label{fig:motivation_balance}
  \includegraphics[scale=.55]{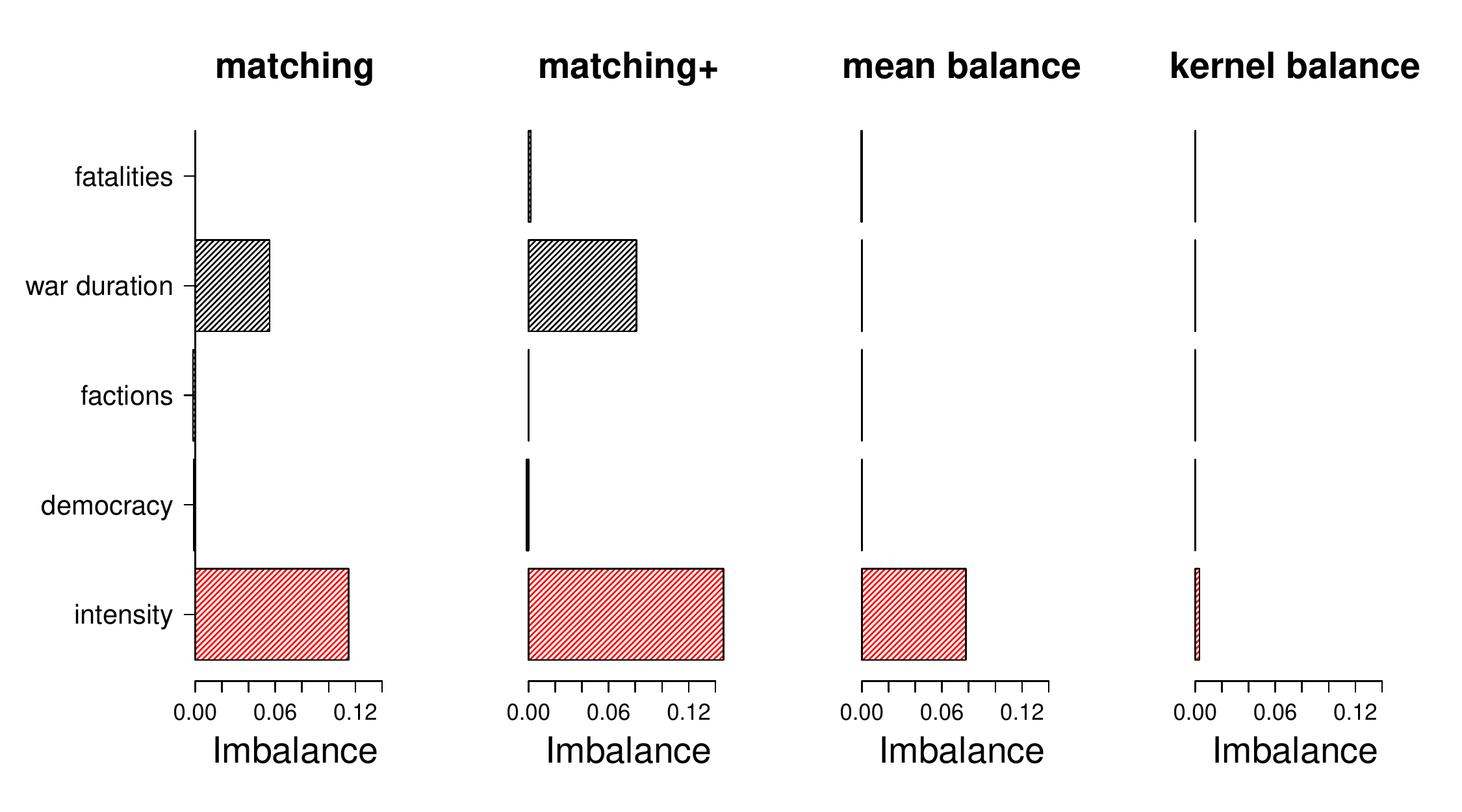}
  \subcaption*{\scriptsize{Mean imbalances on included covariates and $intensity=\frac{fatalities}{war\,duration}$, which determines  both assignment of the treatment ($peacekeeping$) and the outcome ($peaceyears$). \emph{Matching:} Mahalanobis distance matching on the original covariates alone leaves a substantial imbalance on $war\,duration$. More problematically, it shows a large imbalance on $intensity$. \emph{Matching+:} Mahalanobis distance matching with squared terms and all pairwise multiplicative worsens imbalance, particularly on $intensity$. \emph{Mean balance}: Entropy balancing on the original covariates achieves essentially perfect mean balance on these, but only a small improvement in balance on $intensity$. \emph{Kernel balance} obtains mean balance on a wide range of smooth functions of the included covariates, obtaining balance $intensity$ despite not including it in the algorithm.}}
\end{figure}

How well do existing techniques achieve equal means for the treated and controls (``mean balance''), both on the original four covariates and on $intensity$, a (non-linear) function of the observables? In Figure \ref{fig:motivation_balance}, the horizontal axis for each plot shows the standardized difference in means between treated and control on each of the covariates, as well as on $intensity$. All results are averaged over 500 simulations with the same data generating process and $N=500$ on each simulation. The first plot (\emph{matching}) shows results for simple Mahalanobis distance matching (with replacement). Imbalance remains somewhat large on $war\,duration$. More troubling, imbalance remains considerable on $intensity$, which was not directly included in the matching procedure.  A careful researcher may realize the need to match on more functions of the covariates, and instead match on the original covariates, their squares, and their pairwise multiplicative interactions. While few researchers go this far in practice, the second plot in figure \ref{fig:motivation_balance} (\emph{matching+}) shows that even this approach would not provide the needed flexibility to produce balance on $intensity$. In fact, balance on both $war\,duration$ and $intensity$ are worsened. In the third plot (\emph{mean balance}), entropy balancing \citep{hainmueller2012entropy} is used to achieve equal means in the original covariates. As expected, this produces excellent balance on the original covariates, but only a modest improvement in balance on $intensity$. Finally, the fourth plot previews results using the method proposed here, (\emph{kernel balance}). Because this method achieves balance on many smooth functions of the included covariates, it achieves vastly improved balance on $intensity$.
 
\begin{figure}[!hbt]
  \centering
  \caption{Biased ATT estimation due to imbalanced function of the covariates}
  \label{fig:motivation_ATT}
    \includegraphics[scale=.7]{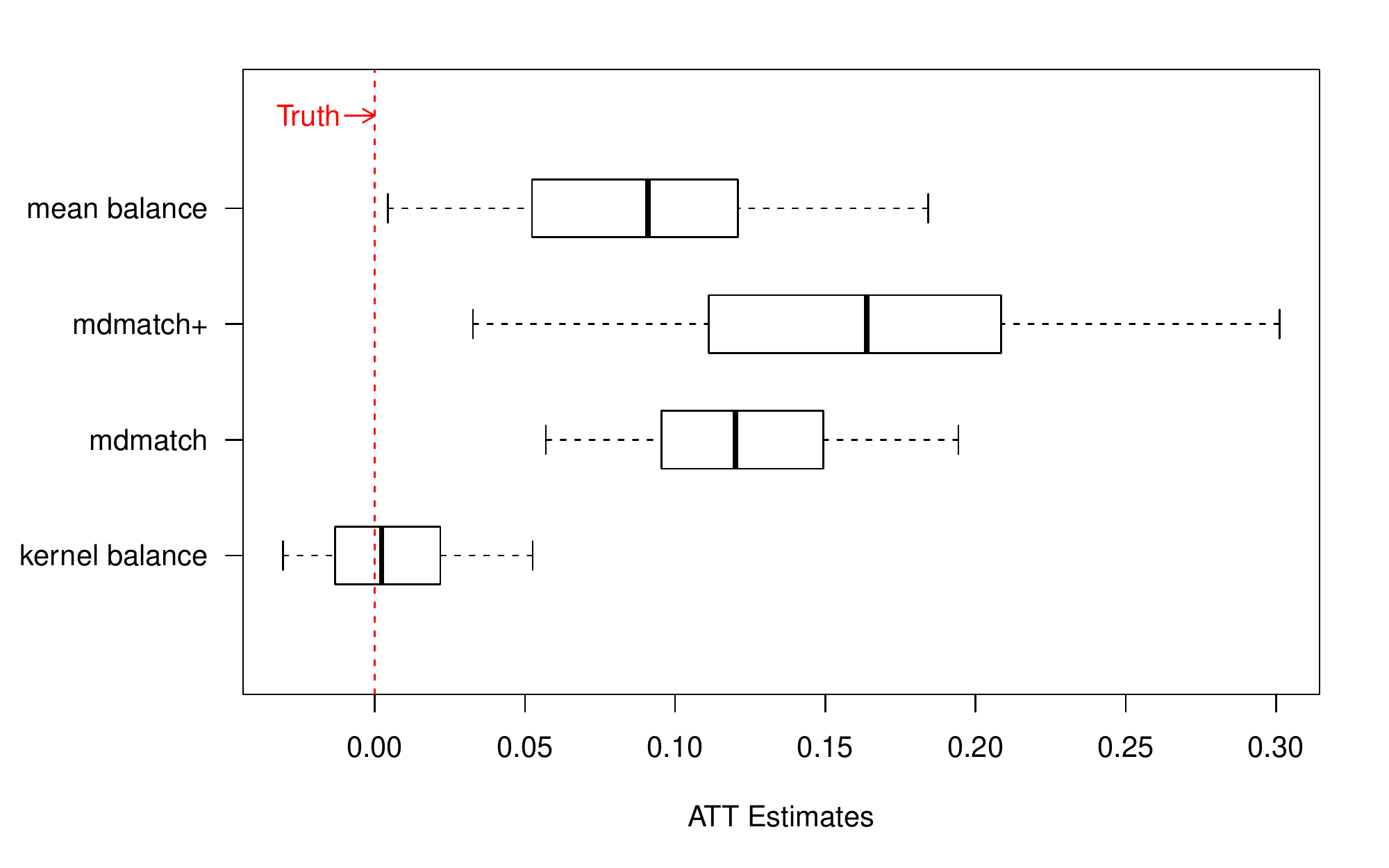}
 \subcaption*{\scriptsize{Boxplot illustrating distribution of average treatment effect on the treated (ATT) estimates in the same example as Figure \ref{fig:motivation_balance} above.  The actual effect is zero $peace\,years$. \emph{Matching}, \emph{matching+}, and \emph{mean balance} all show large biases because the control samples chosen by these procedures include higher $intensity$ conflicts than the treated sample, even though $intensity$ is entirely a function of observables. Since $intensity$ influences the outcome, $peace\,years$, the treated and control samples thus differ regardless of any treatment effect.  By contrast, \emph{kernel balance} is approximately unbiased, as it achieves balance on a large space of smooth functions of the covariates.}}
\end{figure}
 
These imbalances are worrying because they lead to biased ATT estimates. When the ATT is estimated by difference in means in the post weighting/matching sample, larged bias occur with the exception of kernel balancing (Figure \ref{fig:motivation_ATT}). The uncertainy in these estimates come from variability in the sample drawn. The reduced variability in ATT estimates for kernel balancing in figure \ref{fig:motivation_ATT}) is noteworthy, resulting from the high degree of balance on outcome-influencing functions of the covariates in each resample. 

This illustration while artificial, motivates the method described here by showing how easy it is for other methods to produced badly biased estimates: a non-linear function of two observed covariates -- even as simple as a ratio -- may influence the potential outcomes, producing scenarios in which existing matching and weighting methods pose risks of large biases. Theoretical background information on the topic in question is very rarely sufficient to ensure the investigator can guess what functions of the observables may impact the outcome. Kernel balancing, described below, provides one principled approach for choosing function of the covariates on which to achieve balance to ensure unbiased estimation in a wide range of plausible scenarios. 

\section{Framework for Kernel Balancing} \label{proposal}
This section sets up the problem of ATT estimation, then describes the main ideas of the kernel balancing approach. Using the Neyman-Rubin potential outcomes framework (see e.g. \citealp{rubin1990application, neyman1923}) let $Y_{1i}$ and $Y_{0i}$ be the treatment- and non-treatment potential outcomes respectively for units $i=1,2,\ldots,N$, and $D_i \in \{0,1\}$ be the treatment assignment for unit $i$ such that $D_i=1$ for treated units and $D_i=0$ for control units. The observed outcome for each unit is thus $Y_i=D_iY_{1i}+(1-D_i)Y_{0i}$. Suppose each unit has a vector of observed covariates, $X_i$, taking values $x \in \mathcal{X}$ where $\mathcal{X}$ is the support, assumed to lie in $\mathbb{R}^P$. These are assumed to be unaffected by the treatment and are thus called ``pre-treatment'' covariates. For all $i$, assume that the random variables $\{Y_{1i},Y_{0i}, X_i,D_i\}$ are independent with common joint density $p(X,Y_1,Y_0,D)$.

We will be interested in the average treatment effect on the treated, $\E[Y_{1i}-Y_{0i}|D_i=1]$. However, when working with samples it will be more direct to consider the \textit{sample average treatment effect on the treated} (SATT), $\frac{1}{N_1} \sum_{i:D_i=1} \left( Y_{1i}-Y_{0i} \right)$, where $N_1$ is the number of treated and the sum is taken over only treated units. This conditions on the sample in hand, though still requires estimating the unobserved mean, $\frac{1}{N_1} \sum_{i:D_i=1} Y_{0i}$. Because the sample is drawn independently from $p(X,Y_1,Y_0,D)$, $\E[SATT]=ATT$, so unbiased estimates for the SATT will be unbiased for the ATT as well.

\subsection{Bias of Difference in Means}
Consider the (unweighted) difference in means estimand $\text{DIM} \equiv \E[Y_i|D_i=1]-\E[Y_i|D_i=0]$, and its sample analog, $\widehat{\text{DIM}} \equiv \frac{1}{N_1}\sum_{i:D_i=1} Y_i - \frac{1}{N_0}\sum_{i:D_i=0} Y_i$ where $N_0$ is the number of control units.

The DIM is unbiased for the SATT (and the ATT) only when the treated and control groups would have the same expected outcome if neither had received the treatment. This allows the average outcome from the non-treated units to proxy for the average non-treatment potential outcome that the treated units would have had, had they not been treated. We can formalize this by decomposing the DIM estimator into the SATT and a bias (see e.g. \citealp{angrist2008mostly}):

\begin{align}\label{decomp}
\frac{1}{N_1}\sum_{i:D_i=1} Y_i - \frac{1}{N_0}\sum_{i:D_i=0} Y_i
&= \frac{1}{N_1} \sum_{i:D_i=1} Y_{1i} -\frac{1}{N_0}\sum_{i:D_i=0} Y_{0i} \\
&= \frac{1}{N_1} \sum_{i:D_i=1} Y_{1i} - \frac{1}{N_1}\sum_{i:D_i=1} Y_{0i} + \frac{1}{N_1}\sum_{i:D_i=1} Y_{0i} - \frac{1}{N_0}\sum_{i:D_i=0} Y_{0i} \\
&= \mbox{SATT} + \mbox{Bias}
\end{align}

It follows that the DIM is unbiased for the SATT simply when $\frac{1}{N_1}\sum_{i:D_i=1} Y_{0i} = \frac{1}{N_0}\sum_{i:D_i=0} Y_{0i}$, which I will refer to as ``mean balance on $Y_{0i}$.'' 

\begin{lemma}[Mean Balance on $Y_{0i}$ Implies Unbiasedness of DIM for SATT]\label{lemma1}
Provided the relevant moments exist, the difference in means (DIM) estimator is unbiased for the SATT if and only if mean balance on $Y_{0i}$ holds, $\frac{1}{N_1}\sum_{i:D_i=1} Y_{0i}=\frac{1}{N_0}\sum_{i:D_i=0} Y_{0i}$. 
\end{lemma}

A final preliminary we need concerns identification assumptions. As for matching, weighting, regresssion, and propensity score adjustment techniques, kernel balancing is only a method of adjusting for observable differences between treated and control units, and this reveals a causal effect on the outcome only when we presume there are no unobserved sources of differences on the potential outcomes between the treated and control group once we do this. We thus requires that treatment assignment is ignorable with respect to the potential outcomes conditionally on the covariates, which I will refer to as conditional ignorability, though it is also sometimes called simply ``ignorability'', ``strong ignorability'', ``no unobserved confounding'', or ``selection on observables''. When identifying treatment effects averaged only over treated units as in the ATT or SATT, this can be weakened slightly to conditional ignorability of the $Y_{0i}$ alone,

\begin{assumption}[Conditional Ignorability for the Non-treatment Outcome]\label{weakSOO}
We say the non-treatment outcome is conditionally ignorable if $$Y_{0i} \indep D_i \; |\; X_i$$ 
\noindent where $Y_{0i}$ is the non-treatment potential outcome, $D_i$ is treatment status, and $X_i$ is a vector of observed, pre-treatment covariates.
\end{assumption}

\subsection{Obtaining Mean Balance on $Y_{0i}$}
Kernel balancing differs from other methods in how it makes use of Assumption \ref{weakSOO} in constructing an estimator. Matching and weighting methods can be understood as an effort to make the distribution of $X_i$ the same for the treated and control (i.e.  multivariate balance), after which a simple difference-in-means would be unbiased for the ATT under Assumption \ref{weakSOO}. By contrast, kernel balancing targets the much simpler goal of obtaining equal means for $Y_{0i}$ for the treated and the control groups (i.e. mean balance), allowing unbiased SATT estimation by Lemma \ref{lemma1}. The \textit{Discussion} section further explores the differences among methods. 

Establishing mean balance -- without observing $Y_{0i}$ for any units (even the controls) -- requires assumptions on the ways in which $X_i$ is allowed to relate to the expectation of $Y_{0i}$. Specifically, assume $X \in \mathbb{R}^P$ is a set of covariates or characteristics satisfying Assumption \ref{weakSOO}, and $\phi(X): \mathbb{R}^P \mapsto \mathbb{R}^Q$, where $Q$ may be (much) larger than $N$, is an expanded set of these characteristicss to be used as a set of basis functions. The specific nature of $\phi(\cdot)$ used in kernel balancing will relate to a choice of kernel, with a Gaussian kernel used in the particular implementation given here. For the moment, the key feature of $\phi(\cdot)$ needed here is that it is a sufficiently rich, non-linear expansion such that $\E[Y_{0_i}|X_i=x]$ can be well fitted as a linear function of $\phi(x)$:

\begin{assumption}[Linearity of Expected Non-treatment Outcome]\label{Y0linear}
We assume that the conditional expectation of $Y_{0i}$ is linear in the expanded features of $X_i$, $\phi(X_i)$, i.e. $\exists$ $\theta \in \mathbb{R}^Q$ and $\phi(\cdot): \mathbb{R}^P \mapsto \mathbb{R}^Q$ s.t. 
\[\E[Y_{0i}|X_i=x]= \phi(x)^{\top}\theta\] 
\end{assumption}

Assumption \ref{Y0linear} is the critical assumption that must be made under this approach. We will soon see that the choice of $\phi(X_i)$ to be used will be a very general one associated with a kernel, with special attention to the case of the Gaussian kernel.  This will allow the function space $\phi(X_i)^{\top}\theta$ to capture all continuous functions as $N \rightarrow \infty$. More importantly, in finite samples, this space can be understood as the smooth and flexible space of functions that can be built by placing (Gaussian) kernels over the observations, rescaling them as needed, and summing them. This is described at length below and particularly in Section \ref{whatthisspaceis}. In addition, potential violations of Assumption \ref{Y0linear} bias the resulting SATT estimate only if the components of $\E[Y_{0i}|X_i]$ not in the span on $\phi(X_i)$ are correlated with treatment $D_i$ (see Appendix \ref{finitesamplemisspec}).  

Next, consider a choice of non-negative weights $w_i$ on the control units that sum to $1$, such that the weighted average vector $\phi(X_i)$ for the controls equals the unweighted average vector $\phi(X_i)$ for the treated, 

\begin{definition}[Mean balance on $\phi(X)$]\label{meanbalphi}
\; We say that $w$ provides mean balance on $\phi(X)$ when:
\[\frac{1}{N_1}\sum_{i:D_i=1} \phi(X_i) = \sum_{i:D_i=0} w_i \phi(X_i)\]
such that  $\sum_i w_i =1$, and $w_i \geq 0$ for all $i$. 
\end{definition}

In practice, many weights can achieve these constraints. In the implementation used here, following \cite{hainmueller2012entropy}, I choose the weights that satisfy these constraints with maximium uniformity as measured by entropy. See Section \ref{implementation} for implementation details.

Once mean balance on $\phi(X)$ is achieved, all linear functions of $\phi(X)$ then have the same mean for the treated and control groups. To see this, note that the assumption that $\E[Y_{0i}|X_i]$ is linear in $\phi(X_i)$ (Assumption \ref{Y0linear}) is equivalent to assuming a noise model for the data of the form $Y_{0i} = \theta^{\top}\phi(X_i)+\epsilon_i$ with no restrictions on $\epsilon$ except that $\E[\epsilon_i|X_i]=0$. We can then represent the sample mean of $Y_{0i}$ for the treated as
\begin{align}
\frac{1}{N_1} \sum_{i:D_i=1} Y_{0i} &= \frac{1}{N_1} \sum_{i:D_i=1} \left\lbrace \phi(X_i)^{\top}\theta + \epsilon_i \right\rbrace \\
&= \theta^{\top}  \frac{1}{N_1} \sum_{i:D_i=1} \phi(X_i) + \frac{1}{N} \sum_{i:D_i=1} \epsilon_i \label{eq:y0t}
\end{align}

\noindent while the sample mean of $Y_{0i}$ for the controls (after weighting) is

\begin{align}
\sum_{i:D_i=0} w_i Y_{0i} &= \sum_{i:D_i=0}  w_i\left\lbrace\phi(X_i)^{\top}\theta + \epsilon_i \right\rbrace\\
&= \theta^{\top}  \sum_{i:D_i=0} w_i \phi(X_i) + \sum_{i:D_i=0} w_i \epsilon_i \label{eq:y0c}
\end{align}

Recall that the expected bias of the SATT is the expected difference between the mean non-treatment potential outcomes for the treated and controls, which we obtain from the difference between expressions \ref{eq:y0t} and \ref{eq:y0c}:
\begin{align}
\E[Bias] &= \E[\theta^{\top}  \frac{1}{N_1} \sum_{i:D_i=1} \phi(X_i) + \frac{1}{N} \sum_{i:D_i=1} \epsilon_i - \theta^{\top}  \sum_{i:D_i=0} w_i \phi(X_i) + \sum_{i:D_i=0} w_i \epsilon_i]
\end{align}  
\noindent Mean balance on $\phi(X)$ in the sample reduces this to 
\begin{align}
\E[Bias] &= \E \left[ \frac{1}{N} \sum_{i:D_i=1} \epsilon_i - \sum_{i:D_i=0} w_i \epsilon_i \right] = 0
\end{align}
\noindent Mean balance on the linear bases $\phi(X_i)$ is thus sufficient for unbiased SATT estimation under Lemma \ref{lemma1}. Note that the coefficients $\theta$ need not be determined. To put this more simply, when the expectation of $Y_{0i}$ is linear in $\phi(X_i)$, equal means on $\phi(X_i)$ leads to equal means on $Y_{0i}$. 

What remains is to obtain mean balance on these features. If one had sufficient knowledge of $\E[Y_{0i}|X_i]$ to select a low-dimensional choice of $\phi(\cdot)$ while being confident that $\E[Y_{0i}|X_i]$ is linear in $\phi(X_i)$, then one could directly seek mean balance on each dimension of $\phi(X_i)$ and be confident that mean balance on $Y_{0i}$ has been achieved. However, the general supposition of this paper is that investigators usually do not have sufficient knowledge of the functional form of $\E[Y_{0i}|X_i]$ to unfailingly choose a low-dimensional $\phi(\cdot)$.  Typically, little is known about this surface, except perhaps its continuity or anticipated smoothness. A very general choice of $\phi(\cdot)$ is thus required, so that the functions $\phi(X_i)^{\top}\theta$ would include most reasonable functions. Yet, this poses a computational challenge: the higher dimensional choice one makes for $\phi()$, the more difficult and less feasible it becomes to find the balancing weights, particularly if the dimension of $\phi()$ rises above $N$. A convenient choice of $\phi()$ that avoids this choice is made possible through the use of kernels, to which I now turn.  

\subsection{Kernels}\label{kernels}
In this section I explain why kernels in general, and the Gaussian kernel in particular, allow us to employ a very general choice of $\phi(\cdot)$ that guarantees linearity of the expected outcome in these features under mild conditions, while also making the balancing problem tractable even though  $dim(\phi(X_i))>>N$ or indeed when $dim(\phi(X_i))$ is infinite-dimensional. For $X_i \in \mathbb{R}^P$, a kernel function, $k(\cdot, \cdot): \mathbb{R}^P \times \mathbb{R}^P \mapsto \mathbb{R}$, takes in covariate vectors from any two observations and produces a single real-valued output interpretable as a measure of similarity between those two vectors. For reasons discussed below, we are interested principally in the Gaussian kernel:

\begin{equation}\label{similarity}
k(X_j,X_i)=e^{-\frac{||X_j-X_i||^2}{2b}}
\end{equation}

\noindent 
Note that $k(X_i,X_j)$ produces values between 0 and 1 interpretable as a (symmetric) similarity measure, achieving a value close to 1 when $X_i$ and $X_j$ are most similar and approaching $0$ as $X_i$ and $X_j$ become dissimilar.  The choice parameter $b$ might be called ``scale'', because it governs how close $X_i$ and $X_j$ must be in a Euclidean sense to be deemed similar. I discuss the choice of $b$ further below. It is common to rescale each covariate prior to computing $k(X_i,X_j)$, dividing by the standard deviation. This ensures results will be invariant to unit-of-measure decisions. 
 
For a kernel that produces a positive semi-definite (PSD) kernel matrix $\mathbf{K}$ with elements $\mathbf{K}_{(i,j)}=k(X_i,X_j)$,  there exists a choice of basis functions $\phi(\cdot)$ such that $\langle \phi(X_i), \phi(X_j) \rangle = k(X_i,X_j)$. This is due to the equivalence between PSD matrices and Gram matrices formed by inner products of vectors: a PSD matrix $\mathbf{K}$ has spectral decomposition $\mathbf{K}=V \Lambda V^{\top}$, and so $k_{i,j}= (\Lambda^{\frac{1}{2}}V_{[\cdot,i]})^{\top}(\Lambda^{\frac{1}{2}}V_{[\cdot,j]})$. Defining $\phi(X_i)=\Lambda^{\frac{1}{2}}V_{[\cdot,i]}$, we obtain $k_{i,j}=\phi(X_i)^{\top}\phi(X_j)$. The generalization of this to potentially infinite-dimensional eigenfunctions is given by Mercer's Theorem \citep{mercer1909functions}.  

This equivalence between a $k(X_i,X_j)$ and an inner-product $\langle \phi(X_i), \phi(X_j) \rangle$ may not at first seem to be a useful relationship, however it is important because as discussed below (Section \ref{balKgetsbalphi}), it will be possible to achieve balance on $\phi(X_i)$ without having to even form $\phi(X_i)$, by instead achieving balance on vectors made up of these inner-products. 

The nature of the $\phi(X)$ depends on the choice of kernel. For example, suppose $X_i=[X_i^{(1)},X_i^{(2)}]$ and we choose the kernel $(1+\langle X_i,X_j \rangle)^2$. This choice of kernel happens to corresponds to $\phi(X)=[1, \sqrt{2} X^{(1)}, \sqrt{2}X^{(2)},X^{(1)}X^{(1)},\sqrt{2}X^{(1)} X^{(2)},X^{(2)}X^{(2)}]$, and one can confirm that $k(X_i,X_j)=\langle \phi(X_i),\phi(X_j) \rangle$ for this choice of kernel and $\phi(\cdot)$. Using the Gaussian kernel, the corresponding $\phi(X)$ is infinite-dimensional. I describe the function space linear in these features in section \ref{whatthisspaceis}. 

Note that this feature space has universal representation property: as $N \rightarrow \infty$, $\phi^{\top}(X)\theta$ can fit any continuous function of $X$ \citep{micchelli2006universal}. This is less reassuring in small samples. However, smoother functions can be fitted with fewer observations, making this an excellent choice to model $\E[Y_{0i}|X_i]$ when little is known about the nature of the relationship except that it is continuous and likely to be smooth. Further justification for the Gaussian choice of kernel and an intuition for the nature of this feature space (and the functions linear in it) is given in section \ref{whatthisspaceis}. 

Let $\mathbf{K}$ be the kernel matrix storing the results of each pairwise application of the kernel, i.e. $\mathbf{K}_{\{i,j\}}=k(X_i,X_j)=\langle \phi(X_i), \phi(X_j) \rangle$. To reduce notation it is useful to order the observations so that the $N_1$ treated units come first, followed by the $N_0$ control units. Then $\mathbf{K}$ can be partitioned into two rectangular matrices, 

\[\mathbf{K} = 
\begin{bmatrix}
    \mathbf{K_t} \\
    \mathbf{K_c}
\end{bmatrix} \]

\noindent where $\mathbf{K_t}$ is $N_1 \times N$ and $\mathbf{K_c}$ is $N_0 \times N$. The average row of $\mathbf{K}$ for the treated can then be written $\frac{1}{N_t}\mathbf{K_t} \mathbf{1}_{N_t}$, while the weighted average row of $\mathbf{K}$ is $\mathbf{K_c} w$ for the $N_0 \times 1$ vector of weights $w$, with weights summing to 1. 

\subsection{Mean balance on $\mathbf{K}$}\label{meanbalanceonK}\label{balKgetsbalphi}
Working  with kernels and constructing the kernel matrix $\mathbf{K}$ pays off because mean balance on $\phi(X)$ is achieved by getting mean balance on the columns of $\mathbf{K}$. Consider a single row of $\mathbf{K}$:
$$k_i = [k(X_i,X_1), k(X_i,X_2),\ldots,k(X_i,X_N)]$$

\noindent which describes each observation not in terms of its original $X$ coordinates but as a vector of $N$ similarities to each of the observations. Similar to mean balancing on $X_i$, kernel balancing then seeks weights that ensure the average $k_i$ of the treated is equal to the weighted mean vector $k_i$ of the controls:

\begin{definition}[Mean balance on $\mathbf{K}$]\label{meanbalK}
\; The weights $w_i$ achieve mean balance on $\mathbf{K}$ when
$$\overline{k_t} = \sum_{i:D=0} w_i k_i$$
such that  $\sum_i w_i =1$, and $w_i \geq 0$ for all $i$, where $\overline{k_t}$ is the average row of $\mathbf{K}$.
\end{definition}

\noindent 

This achieves mean balance on the corresponding $\phi(X_i)$ regardless of the dimensionality of the feature expansion.
 
\begin{proposition}[Balance in $\mathbf{K}$ implies balance in $\phi(X)$\\]\label{prop:PhiBalance}
Let the mean row of $\mathbf{K}$ among the treated units be given by $\overline{k_t}=\frac{1}{N_t}\mathbf{K_t} \mathbf{1}_{N_t}$ and the weighted mean row of $\mathbf{K}$ among the controls given by $\mathbf{K_c}w$. If $\overline{k_t}=\mathbf{K_c}w$, then $\overline{\phi_t}=\overline{\phi_c}$ where $\overline{\phi_t}=\frac{1}{N_t}\sum_{D_i=1}\phi(X_i)$ and $\overline{\phi_c}=\sum_{D_i=0}\phi(X_i)w_i$.
\end{proposition}

Proposition \ref{prop:PhiBalance} implies that the treated and control groups have the same mean on each dimension of $\phi(X)$ when the rows of $\mathbf{K}$ for the treated and control have the same means, regardless of the dimensionality of $\phi(\cdot)$.  Proof is given in the appendix. 

Finally, the weights $w_i$ that produce mean balance on $\mathbf{K}$ in a finite sample can be used in a difference in means estimation. The main result can now be stated:  
\begin{theorem}[Unbiasedness of Weighted Difference in Means for the SATT]\label{dimwunbiased}
Consider the weighted difference in means estimator, $$\widehat{DIM}_w = \frac{1}{N}\sum_{i:D_i=1} Y_i - \sum_{i:D_i=0} w_i Y_i$$
$$\mbox{such that }
\overline{k_t} = \sum_{i:D=0} w_i k_i, \; \sum_i w_i =1 \; \mbox{and } w_i>0$$
Under assumptions of conditional ignorability for the non-treatment outcome (Assumption \ref{weakSOO}) and linearity of $\E[Y_{0i}|X_i]$ in $\phi(X_i)$ (Assumption \ref{Y0linear}), $\widehat{DIM}_w$ is unbiased for the sample average treatment effect on the treated (SATT) and the (population) ATT.
\end{theorem}

The proof is given in the appendix (\ref{proofunbiased}), though the intuition is simple and helps to summarize the approach: mean balance in $\mathbf{K}$ gives mean balance in $\phi(X)$, which produces mean balance for functions linear in $\phi(X)$, including the conditional expectation of $Y_{i0}$. There I also describe the bias under conditions in which $\E[Y_{0i}|X_i]$ is not fully linear in $\phi(X)$, showing that bias is introduced only when the component of the regression surface not linear in $\phi(X)$ is correlated with treatment assignment.  
 
\subsection{Implementation}\label{implementation}

What remains is to choose the weights, $w_i$ that obtain mean balance on $\mathbf{K}$. We have great flexibliity in the choice of weights, and in particular, a measure of divergence from uniform weights we wish to keep at a minimum subject to achieving the balance constraints in (Definition \ref{meanbalK}). Appendix \ref{entropydetails} describes implementation options consistent with the approach outlined here, and the particular choice implemented in the package \texttt{kbal}, which maximizes the entropy measure, $\sum_i w_i log(w_i)$, as suggested by \cite{hainmueller2012entropy}. 

Another practical concern is that since the columns of $\mathbf{K}$ are highly colinear, it is preferable to work with a lower-rank approximation. A natural choice would be the rank-$r$ approximation, $\Kr$, closest to $\mathbf{K}$ in the Frobenius norm, i.e. minimizing 
$$||\mathbf{K}-\widetilde{\mathbf{K}}^{(r)}||_\mathcal{F}  = \sqrt{\sum_{i=1}^N \sum_{j=1}^N  |\mathbf{K}_{i,j} - \Kr_{i,j}|^2}$$ 
 
More directly, recall that our aim in achieving mean balance on $\mathbf{K}$ is to ensure balance on linear combinations such as $\mathbf{K}u$ for some $N \times 1$ vector $u$.  In choosing a rank $r$ approximation, we want to ensure that for any $c$ of a particular size $||c||$, $\Kr c$ and $\mathbf{K} c$ cannot be too far apart. Thus, it is desirable to minimize the operator 2-norm:

$$ ||\mathbf{K}-\Kr||_2 = sup \; \frac{||\mathbf{K}c - \Kr c||_2}{||c||_2} $$

Fortunately, among all rank $r$ matrices, the choice of $\Kr$ minimizing both $||\mathbf{K}-\Kr||_2$ and $||\mathbf{K}-\Kr||_{\mathcal{F}}$ is given by principal components analysis (PCA; \citealp{eckart1936approximation}). Since PCA constructs $\Kr$ as a linear combination of the first $r$ principal components of $\mathbf{K}$, we can directly seek mean balance on the those principal components alone.  What remains is the choice of $r$, which can be chosen to minimize the resulting imbalance in $\mathbf{K}$. Details are provided in \ref{entropydetails}. I now turn to establishing this link between balance on $\mathbf{K}$ and equality of kernel estimates of multivariate density for the treated and controls.

\subsection{Smoothed multivariate balance}\label{section:mvb}
The principle motivation for kernel balancing is as a reliable and hands-off method for estimation of the ATT (or ATC or ATE, see Section \ref{otherqois}) by obtaining mean balance on $Y_{0}$ as described above, under reasonable assumptions on $\E[Y_{0i}|X_i]$. 

However, how does the procedure relate to methods such as matching that seek to make the multivariate density of the controls approximately equal to that of the treated? The use of kernels for the choice of $\phi(X_i)$ above produces a very useful equivalence: kernel balancing using kernel $k(\cdot,\cdot)$ implies that for a kernel density estimator also using kernel $k$, the multivariate density of the covariates so estimated is equal for the treated and control groups at all locations in the dataset. It thus also achieves in a finite sample the goal of ``multivariate balance'' normally targeted by matching and weighting procedures, but only insofar as those densities are estimated using the same kernel.

These multivariate density estimators may not be satisfactory density estimators as such, particularly in high-dimensional data. However, methods seeking multivariate density balance can typically only hope to achieve or verify that balance with respect to some density measure anyhow, making this is a very useful equivalence. As a corrolary, a researcher seeking multivariate density balance could first commit to a kernel smoother she would would be willing to use to estimate the multivariate density in each group, after which kernel balancing produces the weights resulting in equality of these estimated densities, when a feasible solution exists. 

\begin{proposition}[Balance in $\mathbf{K}$ implies equality of smoothed multivariate densities\\] \label{prop:densityequalization}
Consider a density estimator for the treated, $\hat{p}_{X|D=1}$ and for the (weighted) controls, $\hat{p}_{X|D=0,w}$, each constructed with kernel $k(\cdot,\cdot)$ of bandwidth $b$ as described below. The choice of weights that ensures mean balance in the kernel matrix $\mathbf{K}$ ensures that $\hat{p}_{X|D=1}=\hat{p}_{X|D=0,w}$ at every position at which an observation is located.
\end{proposition}

Proof of proposition \ref{prop:densityequalization} is given in the appendix. Here I briefly build an intuition for this result, as it leads to further insights. First, the typical Parzen-Rosenblatt window approach estimates a density function according to:
\begin{align}
\hat{p}(x) &= \frac{1}{N\sqrt{2\pi b}}\sum_{i=1}^{N}k(x,X_i) \label{parzen}
\end{align}

\noindent for kernel function $k(\cdot,\cdot)$ with bandwidth $b$.

The Gaussian kernel is among the most commonly used for this task. While typically considered in a univariate context, expression \ref{parzen} utilizing a Gaussian kernel generalizes to a multivariate density estimator based on Euclidean distances. Such density estimators are intuitively understandable as a process of placing a multivariate Gaussian kernel over each observation's location in $\mathbb{R}^P$, then summing them into a single surface and rescaling, providing a density estimate at each location. 

The link between obtaining mean balance on $Y_{0i}$ and obtaining multivariate density balancing emerges from the fact that both are manipulations of the superpositions of kernels placed over each observation. For a sample consisting of $X_1,\ldots,X_N$, construction of the kernel matrix $\mathbf{K}$ using the Gaussian kernel and right-multiplying it by a column vector, $\frac{1}{N\sqrt{2\pi b}}$, produces values numerically equal to first constructing such an estimator based on all the observations represented in the columns of $\mathbf{K}$, then evaluating the resulting density estimates \emph{at all the positions represented by the rows of $\mathbf{K}$}. To see this, consider that the value of $\mathbf{K}a$ at a given point $X_j$ is $\sum_i a_i k(X_i,X_j)$. Note that $k(X_i,X_j)$ is the value that would be obtained by placing a Gaussian over $X_i$ and evaluating its height at $X_j$. Thus $\sum_i a_i k(X_i,X_j)$ is the value that would be obtained by placing a Gaussian kernel over each observation, $X_i$, and evaluating the height of the resulting summated surface at $X_j$. Similarly, the expression $\frac{1}{N_1 \sqrt{2\pi b}}\mathbf{K_t}^{\top}\mathbf{1}_{N_1}$ where $\mathbf{1}_{N_1}$ is a $N_1$-vector of ones thus returns a vector of estimates for the density of the treated, as measured at all observations. Finally, $\frac{1}{N_0 \sqrt{2\pi b}}\mathbf{K_c}^{\top}\mathbf{1}_{N_0}$ returns estimates for the density of the control units at every datapoint in the sample, and $\frac{1}{\sqrt{2\pi b}} \mathbf{K_c}^{\top} w$ gives the $w$-weighted density of the controls, again as measured at every observation. 

If we take these estimates as reasonable measures of density, we would like to choose the weights such that the weighted density of the controls equals that of the treated, at every observation. Proposition \ref{prop:densityequalization} states that the choice of $w$ found by kernel balancing to achieve $\mathbf{K}_c^{\top}w=\frac{1}{N_1} \mathbf{K}_T^{\top}\mathbf{1}_{N_1}$ is exactly the choice that equalizes these smoothed density estimates for the treated and weighted controls at every point in the dataset. Proof is given in the appendix.

Finally, it is useful to introduce a measure of imbalance that can be used prior to or after reweighting. As shown in Appendix \ref{L1equivalence}, a p-norm on the imbalance in terms of $\mathbf{K}$,  $||\overline{k_t} - \sum_{i:D=0} w_i k_i ||_p$ is proportional to the same  p-norm over the pointwise ``gaps'' between the multivariate densities of the treated and controls, as measured by the corresponding kernel density estimator at every point in the sample. Indeed, this is the same norm that is minimized during the choice of $r$ (Appendix \ref{entropydetails}). Here, I report the $L_1$ norm, $\frac{1}{2}||\frac{1}{N_1 \sqrt{2\pi b}}\mathbf{K_t}^{\top} \mathbf{1}_{N_1} -\frac{1}{\sqrt{2\pi b}}\mathbf{K_c}^{\top}w ||_1 = \frac{1}{2}||\hat{p}_{D=1}(\mathbf{X})-\hat{p}_{w,D=0}(\mathbf{X})||$
since it is naturally interpertable as an average of the pointwise gaps between the density of the treated and control at every observation. This is analogous to the $L_1$ norm proposed by \citep{CEMjasa} for use with coarsened exacted matching, but here does not require coarsening the covariates into discrete bins as proposed there.

\begin{figure}[!hbt]
  \centering
  \caption{Density Equalizing Property of the \emph{kbal} Weights}
  \label{fig:densitybalancing}
    \includegraphics[width=.45\textwidth]{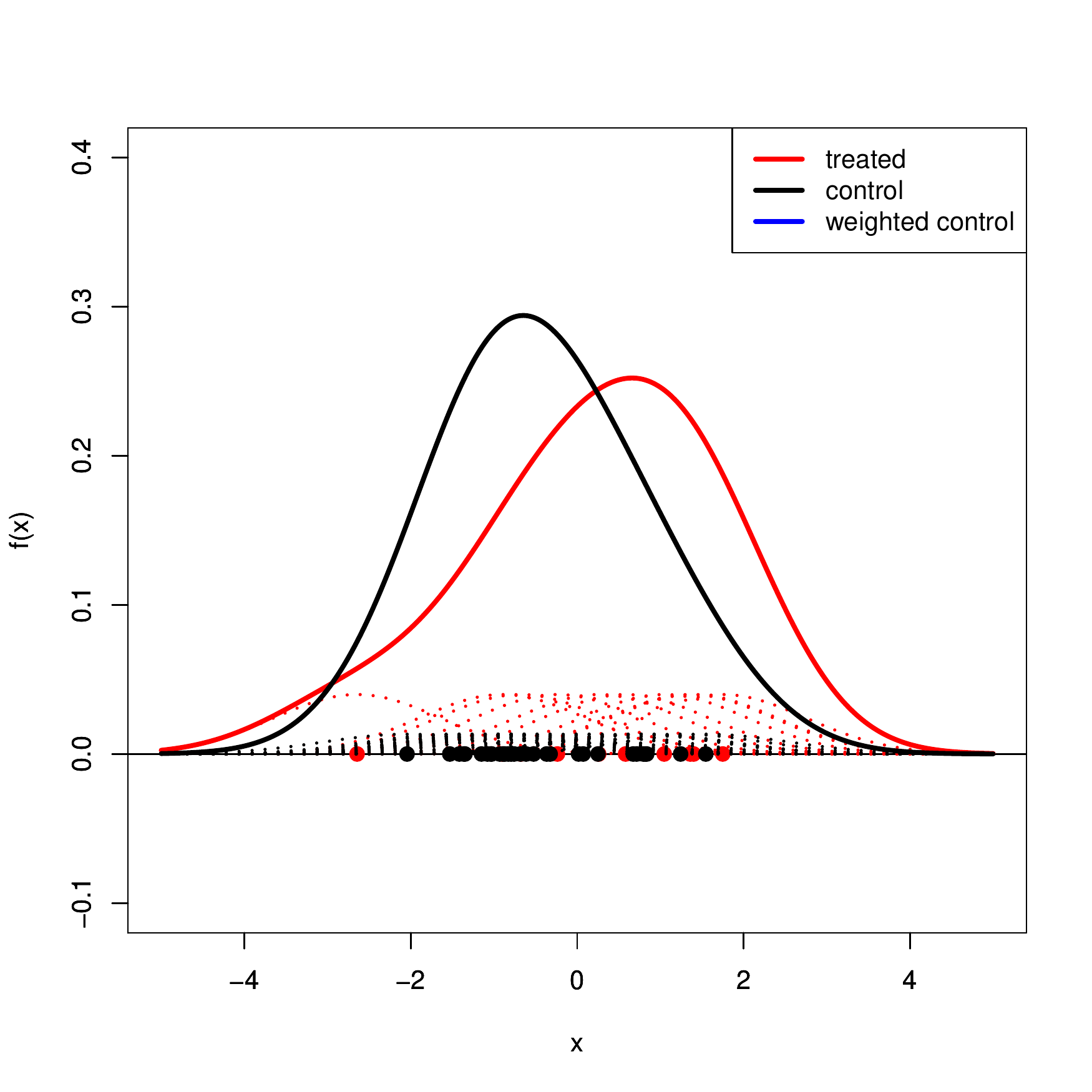}
        \includegraphics[width=.45\textwidth]{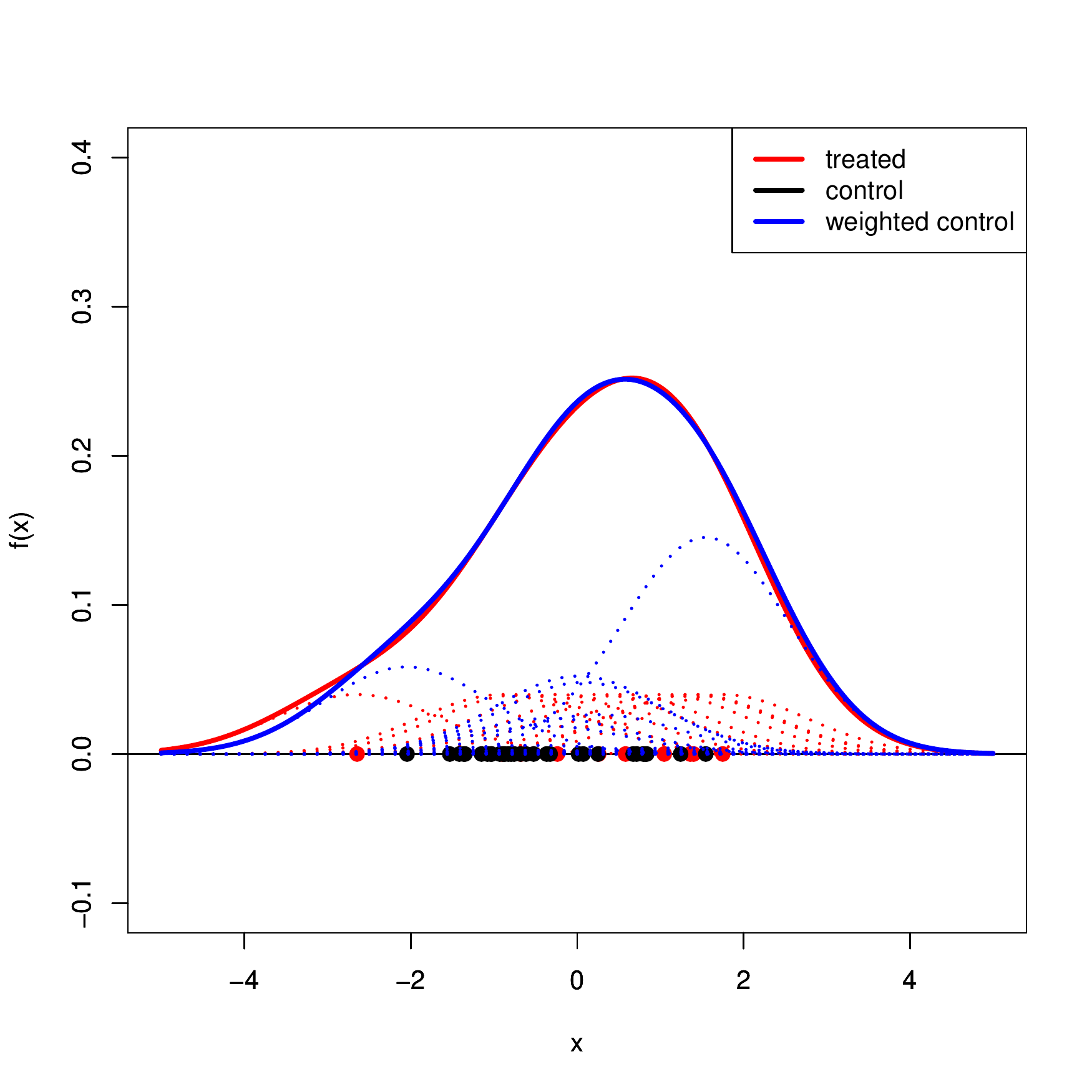}
    \subcaption*{\scriptsize{\emph{Left:} Density estimates for treated and (unweighted) controls.  Red dots show the location of 10 treated units. Dashed lines show the appropriately scaled Gaussian over each observation, which sum to form the density estimator for the treated (red line) and control (black line). The $L_1$ imbalance is measured to be 0.32. \emph{Right:} Weights chosen by kernel balancing effectively rescale the height of the Gaussian over each control observation (dashed blue lines). The new density estimate for the weighted controls (solid blue line) now closely matches the density of the treated at each point. The $L_1$ imbalance is now measured to be $0.002$}}
\end{figure}

Figure \ref{fig:densitybalancing} provides a graphical illustration of the density-equalizing property of the kernel balancing weights for a one-dimensional problem. This density equalizing view connects kernel balancing more directly to other approaches such as matching, but it is important to remember that it is mean balance in $Y_{0i}$ that is essential for unbiasedness, and which kernel balancing targets. Kernel balancing only equalizes the densities \textit{as they are estimated} by the smoothing action of the selected kernel. In some cases a density estimate constructed in this way would not be a natural one, for example when $X$ is a categorical variable or has sharp bounds. Nevertheless, this approach will apply the same smoothing estimator to the treated and to the control. 

\section{Discussion}\label{discussion}

Having described the basic logic and procedure for kernel balancing, I now remark on its relationship to existing procedures, some additional properties and implications of this approach, and further implementation details. 

\subsection{Relation to Existing Approaches}

Here, I compare kernel balancing to matching, covariate balancing weights, and propensity score methods. Like kernel balancing, each of these begins with an ignorability assumption (Assumption $\ref{weakSOO}$). However, these methods exploit Assumption \ref{weakSOO} to make causal inferences through the more difficult estimation route of seeking  multivariate balance rather than merely balance on $Y_0$. I also briefly contrast the approach to the more traditional strategy of simply fitting an outcome model in a suitable space of functions.
 
\subsubsection*{Matching} 
Under conditional ignorability as defined in Assumption \ref{weakSOO}, treatment assignment is independent of potential outcomes within each stratum of $X$. The most natural way to exploit this for estimating the SATT is to perform this conditioning on $X$ very literally: take difference-in-means estimates of the treatment effect within each stratum of $X$, then average these together over the empirical distribution of $X$ for the treated. Subclassification and exact matching estimators do this. However, conditioning on $X$ in this way is impractical or impossible when $X$ is continuous or contains indicators for many categories, since we cannot literally compute differences for each stratum of $X$. 

Matching approaches (e.g. \citealp{rubin1973matching}) mimic this conditioning, taking each treated unit in turn, finding the nearest one or several control units, and retaining only these control units in the sample (typically with replacement). A difference-in-means on the outcomes in the resulting matched data is the same as an average over the differences within each pairing. The method works when multivariate balance is achieved through the matching procedure, i.e. the distribution of $X$ for the control units becomes the same as the distribution for the treated units. The non-parametric nature of matching is appealing as a multivariate balancing technique, but its accuracy is limited by the problem of matching discrepancies. Specifically, in a given pairing, the treated unit may be systematically different on $X$ than the control unit(s) it is paired with when exact matches cannot be found. Thus the conditioning on $X$ is incomplete, and the distribution of $X$ for the treated and controls are not identical. The resulting bias in (S)ATT estimates dissipates only very slowly as $N$ increases, and in general the resulting estimates are not $\sqrt{N}$-consistent (\citealp{abadie2006large}). 

To minimize bias due to remaining matching discrepancies, investigators are instructed to attempt different matching specifications and procedures until they achieve satisfactory multivariate balance (see e.g. \citealp{stuart2010matching}). However in practice, tests for this balance are usually limited to univariate tests comparing the marginal distribution of each covariate under treatment and control. In short, the goal of matching is to align the multivariate distribution of covariates for the control units with that of the treated, but matching discrepancies can prevent this from occurring, and the tools used to test for this multivariate balance are incomplete. As the motivating example in Section \ref{motivation} illustrates, matching can thus fail to obtain sufficient similarity of distributions, even when investigators attempt to match on higher-order terms.

\subsubsection*{Covariate Balancing Weights}
Another category of methods for multivariate balancing is covariate balancing weighting techniques that use probability-like weights on the control units to achieve a set of prescribed moment conditions on the distribution of the covariates (e.g. univariate means and variances). Examples from the causal inference literature include entropy balancing \citep{hainmueller2012entropy} and the covariate balancing propensity score \citep{imai2014covariate}, with a number of similar procedures emerging from the survey sampling literature, such as raking \citep{raking}. Once these moment conditions are satisfied, it is assumed that the multivariate densities for the treated and control are alike in all important respects. These weights can be used in a difference in means estimation or other procedure. The upside of this procedure over matching is that the prescribed moments of the control distribution can often be made exactly equal to those of the treated, avoiding the matching discrepancy problem.  The downside is that it loses the non-parametric quality of matching, providing balance only on enumerated moments. It is generally not possible to know what moments of the distribution must be balanced to ensure unbiasedness, because we do not know which functions of the covariates might influence the (non-treatment) outcome. Kernel balancing can be understood as an extension to these covariate balancing weighting methods that solves this problem by ensuring balance on a large class of functions of the covariates automatically.

\subsubsection*{Propensity Score Weighting}
Propensity score methods such as inverse propensity score weighting can similarly be understood as an attempt to find the weights that make the distribution of the covariates for the controls and treated similar (in expectation), but through adjusting for estimated treatment probabilities.

For purposes of ATT estimation, the stabilized inverse propensity score weights applied only to the control units would be $w_{IPW} = \frac{p(D_i)}{p(D_i|X_i)}\frac{1-p(D_i|X_i)}{1-p(D_i)}$. Appendix \ref{appendix:ipw} shows how these weights can be derived as those that transport the distribution of the controls to match that of the treated during ATT estimation. As also shown there, these weights can be rewritten via Bayes rule as the ratio of class densities for the treated and controls,
\begin{align}
w_{IPW} &= \frac{p(x|D_i=1)}{p(x|D_i=0)} \label{IPW}
\end{align}

Written in this way, it becomes clear that whenever the class densities are equal for the two groups, the IPW weights would have to remain constant at 1. This makes sense, since two classes with identical multivariate distributions would indeed be indistinguishable, producing constant propensity scores under a generative model for the probability of taking the treatment. Given the multivariate balancing property discussed above, kernel balancing weights achieve precisely this equality of class densities, insofar as multivariate density is estimated by the corresponding kernel density estimator (Section \ref{section:mvb}). This provides an intimate relationship between kernel balancing and inverse propensity score weighting: inverse propensity score weights become constant (and thus unnecessary) in a sample that has been weighted by kernel balancing already, but only when the corresponding kernel density estimator is used. Yet, kernel balancing does not explicitly model a propensity score, nor even restrict it to a particular function space.

\subsubsection*{Comparison to Outcome Models}
An alternative and common estimation route is simply to regress the observed $Y_i$ on some (possibly augmented) set of covariates $X_i$ and treatment $D_i$.  

Kernel balancing assumes the existence of (but does not estimate) an outcome model $\E[Y_i|X_i]=\phi(X_i)^{\top}\theta= k_i^{\top } \bc $. Rather than fitting such a model or otherwise utilizing the outcome data, kernel balancing uses this assumed existence as a device for determining what basis functions need to have the same mean for the treated and control groups in order to ensure that the $\E[Y_{0i}|D_i=1]=\E[Y_{1i}|D_i=0]$. Indeed, once two samples have equal means on $\phi(X)$ (or equivalently, $k_i$), no outcome model need be employed to estimate the ATT -- difference in means is sufficient.  That said, double-robust estimation options that further include covariates and an outcome model on the weighted data can be considered as well (see \citealp{ebaldouble} for double-robustness and optimal efficiency results in the special case where entropy balancing is used to achieve mean balance directly on $X$ and $Y_{0i}$ is assumed linear in $X$).

Two important distinctions can be made between assuming an outcome model for purposes of choosing ``what to balance on'' versus fitting an outcome model, either to directly estimate an effect of treatment or to predict $\E[Y_{0i}|X_i]$ at each location $X_i$ where a treated unit found in order to impute the missing potential outcome. The first is that kernel balancing works regardless of the value of $\theta$ or $\bc$, and we do not need to rely on the accuracy of estimates for these quantities in a finite sample. We need only that such a model exists, and even then, violations of the model are bias-inducing only in certain cases (see Appendix \ref{finitesamplemisspec}). 

Second and more importantly, performing a weighting approach whose justification is rooted in a choice of outcome models is not  equivalent to using the outcome model alone, because the former changes the distribution of (in this case) the control group to be more similar to that of the treated prior to estimation of an effect. Such a ``pre-processing'' approach \citep{ho2007matching} is very helpful: once the treated and control groups are made similar in their characteristics through this reweighting, the investigator no longer requires heroic modeling assumptions to bridge the gap between treated and control units that may lie far apart in the covariate space. This point is not a trivial one, and the results in the empirical example below (Section \ref{empirical}) provide one dramatic illustration. There, treated and control distributions differ radically in their distribution of the covariates and simply using an outcome model with covariates and a treatment indicator does a very poor job of estimating how differences in the covariates should be used to implicitly adjust the outcomes over such a wide range of $X$. As a result it estimates small or even negative effects of a job training program on income, while matching and weighting estimators reveal a positive effect that closely matches the experimental result.

\subsection{Uncertainty Estimation}
In most contexts, investigators require a measure of uncertainty such as a standard error or confidence interval around their effect estimates. With matching estimators, a common approach is to ignore the uncertainty due to the matching procedure itself. For example \cite{ho2007matching} argue that since variance estimators for parametric models typically take the data as fixed anyway, when data are pre-processed by a matching procedure, the matched dataset can be taken as fixed for subsequent analyses as well. Thus, the variance can be esimated for parametric outcome models on the matched data in the usual way, i.e. by applying weights that reflect which control units are dropped or multiply used to the outcome model of interest and computing the associated standard errors. Similarly, weighting estimators such as entropy balancing may also take this pre-processing view and treat the resulting weights as fixed (\citealp{hainmueller2012entropy}) for purposes of computing uncertainty estimates in subsequent analyses. 

In contrast, \cite{abadie2008failure} consider the uncertainty due to the matching process, noting that the bootstrap fails in this case due to the ``extreme non-smoothness'' of matching estimators. \cite{abadie2006large} develop asymptotic standard errors that do account for uncertainty in the matching procedure. Others have argued that an m-out-of-n bootstrap may be appropriate (see \citealp{politis1994large}).

One benefit of kernel balancing and other weighting methods is that, because the weights are continuous and observations are not wholly dropped as in matching, the simple bootstrap is likely to be valid. While further work is needed on more computationally attractive alternatives, boostrapping the entire procedure of selecting weights by kernel balancing then then estimating the subsequent treatment effect is likely an appropriate choice for users who wish to incorporate uncertainty from the weight selection into the final estimates. 

\subsection{Intuition for $\phi(X_i)$ and function space $\phi(X_i)^{\top}\theta$}\label{whatthisspaceis}

A key assumption of the method is that $\E[Y_{0i}|X_i]$ can be well fitted by $\phi(X_i)\theta$ (Assumption \ref{Y0linear}), where $\phi(\cdot)$ is determined by a particular choice of kernel. In this implementation, I focus on the Gaussian kernel, and so it is useful to understand what this function space looks for this choice.  

This function space is the same Reproducing Kernel Hilbert space of functions used by numerous regression and classification methods employing a Gaussian kernel, including kernel ridge regression, support vector machines, and Gaussian processes. Since the choice of $\phi(X_i)$ implied by the Gaussian kernel is infinite-dimensional, it may seem difficult to imagine what this function space looks like. In fact the choice of $\phi(X)$ such that $\langle \phi(X_i),\phi(X_j) \rangle = k(X_i,X_j)$ is not unique. One valid choice for $\phi(X)$ in the case of the Gaussian kernel is the sequence given by $\left\lbrace \sqrt{\frac{2^d}{d!}} exp(-X_i^2)(X_i)^d  \right\rbrace$ for $d=0,1,...,\infty$ (see Appendix \ref{phivector}). More usefully, however, one can think of $\phi(X_i)$ as simply $k(X_i, \cdot)$ (the ``canonical feature mapping''), and the functions linear in $\phi(X_i)$ as those built from the superposition of Gaussians placed over each observation and arbitrarily rescaled. That is, in the original covariates space $\mathbb{R}^P$, suppose we place a $p$-dimensional Gaussian kernels over each observation in the dataset, rescale each of these by a scalar $c_i$, then sum these rescaled Gaussians to form a single surface. By varying the values of $c_i$, an enormous variety of smooth functions can be formed in this way, approximating a wide variety of non-linear functions of the covariates. This view is described and illustrated at length in \cite{krlspaper}, where this function space is used to model highly non-linear but smooth functions.

This space of functions is appealing because while making no assumptions of linearity or additivity in $X$, it is generally reasonable to assume that the conditional expectation of $Y_{0i}$ is continuous and relatively smooth over $\mathcal{X}$.
As noted above, this feature space has universal representation property, such that as $N \rightarrow \infty$, $\phi^{\top}(X)\theta$ can fit any continuous function of $X$ \citep{micchelli2006universal}. While less reassuring in small samples, the superposition of Gaussians views makes clear that smoother functions can be fitted with fewer observations, making this an excellent choice to model $\E[Y_{0i}|X_i]$ when little is known about the nature of the relationship except that it is continuous and likely to be smooth. Accordingly, the Gaussian kernel is the ``workhorse'' choice for many kernelized regression and classification models. 

By achieving equal expectations for the treated and control on the columns of $\mathbf{K}$, kernel balancing thus ensures that the many smooth functions that can be built by the superposition of Gaussians will have the same mean for the treated and control group. It is thus suitable when we have little knowledge of the shape of $\E[Y_{0i}|X_i]$ but believe it is well approximated in such a flexible functions space.

\subsection{Detailed Choice of Kernel}\label{sigmachoice}
Using the kernel as defined by \ref{similarity} for some choice of $b$, any continuous function $\E[Y_{0i}|X_i]$ can be consistently estimated by functions linear in $\phi(X_i)$. However, some kernel choices work better than others in a sample of limited size. Accordingly, in machine learning applications utilizing kernels, it is common to consider details of the kernel definition that may improve the ability to fit the target function linearly in $\phi(X_i)$ (or equivalently, the columns of $\mathbf{K}$) when the sample size is limited. Here we consider the scaling and rotation of $X$, and the choice of $b$.

The first consideration of this type is how $X$ is scaled and rotated. If some variables in $X_i$ have variances orders of magnitude larger than others, the columns of $\mathbf{K}$ will reflect mostly distances on the largest variables, providing little information on distances among the smaller variables. This is unproblematic as the sample size grows to infinity -- the superposition of Gaussians will still allow flexible modeling of the target functions in the limit. But in a small sample, it limits the quality of fit. It is thus common to utilize a Gaussian kernel that computes the Euclidean distance over variables that have been rescaled to have the same variance. This also has the benefit of making the results invariant to any unit-of-measure decisions. Kernel balancing utilizes this approach. Beyond this, some investigators also wish to make the results invariant to rotation, utilizing a Mahalanobis distance rather than Euclidean distance in the Gaussian kernel. This is left as an option in kernel balancing as implemented here.  

Second, $b$ must be chosen. Since mean balance on $Y_{0i}$ is the primary goal, not density estimation or equalization, the choice of the kernel and $b$ should be made accordingly. While it is tempting to think of $b$ as the usual bandwidth that must be carefully selected in density estimation procedures, here it is much more important to choose $b$ according to how it effects mean balance in $Y_{0i}$. To this end, the choice of parameter $b$ is a feature-extraction decision that determines the construction of $\phi(X_i)$ and thus $\mathbf{K}$. It determines how close two points $X_i$ and $X_j$ need to be in order to have highly similar rows $k_i$ and $k_j$. This implies a bias-variance tradeoff. If $b$ is too large, mean balance is easier to achieve and the weights will have low variance, the resulting balance is less precise (and the corresponding smoothed densities more ``blurred''). If $b$ is too small, $\mathbf{K}$ will approximate the identity matrix, and each row $k_i$ will be nearly linearly independent. In this case, the algorithm will not converge as balance cannot be attained. (The possibility of trimming away treated units that are difficult to match under small $b$ is discussed in Appendix \ref{Trimming}). 

Fortunately, in many cases balance is achievable across a wide range of $b$ values, and estimated SATTs are stable across a wide range. While lower values of $b$ are generally preferable, they risk higher variance, potentially placing large weights on a small proportion of the controls. For an easily interpretable metric, I propose the quantity \textit{min90}, which is the minimum number of control units that are required to account for 90\% of the total weight among the controls.  For example, if \textit{min90}=20, 90\% of the total weight of the controls comes from just the 20 most heavily-weighted observations. This gives the user a sense of how many control units are effectively being used. The empirical example below shows how this can be used.

One reasonable choice that may be a useful reporting standard would be to use $b=dim(X)$, while showing results at other choices for robustness. The square of $\E[||X_i-X_j||],$ used in the exponent of the kernel calculation (\ref{similarity}) scales with $dim(X)$. Choosing $b$ proportional to $dim(X)$ thus ensures a relatively sound scaling of the data, such that some observations appear to be closer together, some further apart, and some in-between, regardless of $dim(X)$. A similar logic has been proposed for regression technique using a Gaussian kernel (see e.g. \citealp{krlspaper,scholkopf2002learning}). The constant of proportionality remains open to debate, but the choice of $b=dim(X)$ has offered very good performance. This is the default value of $b$ used here, though clearly further work is needed on this point.  Investigators may wish to present their results across a range of $b$ values to ensure this choice is not consequential in a given application.  Shoudl the results vary across $b$ values, inspecting $L_1$ and the concentration of weights (e.g. through $min90$) can be helpful for determining an appropriate value.
 
\subsection{Other Quantities: ATE, ATC}\label{otherqois}
I have focused on the ATT for simplicity of exposition, but with minor adjustment this method can also be used to identify the average treatment effect on the controls (ATC) and the average treatment effect on the treated. 

To estimate the ATC, informally speaking we wish to ``move the treated to the control locations'' instead of the other way around. Accordingly, we instead seek weights on the treated units such that the weighted sum of $k_i$ among the treated equals the (unweighted) average among the controls. That is rather than seeking the non-negative weights summing to one such that 
$\overline{k_t} = \sum_{i:D=0} w_i k_i$, we would instead seek the weights:
$$ \overline{k_c} = \sum_{i:D=1} w_i k_i, \; \sum_i w_i =1 \; \mbox{and } w_i>0$$

\noindent where $\overline{k_c}$ is the empirical average $k_i$ taken over the controls only. 

Similarly, for the ATE the goal is to transport both the treated and control to the same location and (more importantly) the same expectation of $Y_{0i}$. Thus we would seek the weights $w^{(1)}_i$ on the treated and $w^{(0)}_i$ on the controls such that 
$$\sum_{i:D=0} w^{(0)}_i k_i = \sum_{i:D=1} w^{(1)}_i k_i = \overline{k}$$
\noindent where $\overline{k}$ is the empirical average of $k_i$ taken over all the observations, treated and control alike.

The \texttt{KBAL} package estimates the ATT by default but optionally estimates the ATC and ATE as well. Note that identification of the ATC requires an assumption analogous to \ref{weakSOO} but for both non-treatment outcomes, specifically $Y_{1i} \indep D_i | X_i$. The ATE requires ignorability with respect to both of the potential outcomes conditionally on $X_i$, $\{Y_{0i},Y_{1i} \} \indep D_i | X_i$.

\section{Example: National Supported Work Demonstration} \label{empirical}
It is useful to know whether kernel balancing accurately recovers average treatment effects in observational data under conditions in which an approximately ``true'' answer is known. This can be approximated using a method and dataset first used by \cite{lalonde1986evaluating} and \cite{dehejia1999causal}, and which has become a routine benchmark for new matching and weighting approaches (e.g. \citealp{diamond2005genetic,CEMjasa,hainmueller2012entropy}).

The aim of these studies is to recover an experimental estimate of the effect of a job training program, the National Supported Work (NSW) program. Following \cite{lalonde1986evaluating}, the treated sample from the experimental study is compared to a control sample drawn from a separate, observational sample. Methods of adjustment are tested to see if they accurately recover the treatment effect despite large observable differences between the control sample and the treated sample. See \citep{diamond2005genetic} for an extensive description of this dataset and the various subsets that have been drawn from it.  Here I use 185 treated units from NSW, originally selected by \cite{dehejia1999causal} for the treated sample. The experimental benchmark for this group of treated units is \$1794, which is computed by difference-in-means in the original experimental data with these 185 treated units. The control sample is drawn from the Panel Study of Income Dynamics (PSID-1), containing 2490 individuals.

The pre-treatment covariates available are age, years of education, real earnings in 1974, real earnings in 1975 and a series of indicator variables: Black, Hispanic, and married. However, as this dataset has now been used many times, it is common practice to use three further variables that are actually non-linear transforms of these: indicators for being unemployed (having income of \$0) in 1974 and 1975, and an indicator for having no highschool degree (fewer than 12 years of education).

As found by \cite{dehejia1999causal}, propensity score matching can be effective in recovering reasonable estimates of the ATT, but these results are highly sensitive to specification choices in constructing the propensity score model \citep{smith2001reconciling}. \cite{diamond2005genetic} use genetic matching to estimate treatment effects with the same treated sample. While matching solutions with the highest degree of balance produced estimates very close to the experimental benchmark, these models included the addition of squared terms and two-way interactions, not to mention the constructed indicators for zero income in 1974 and 1975.  Similarly, entropy balancing \cite{hainmueller2012entropy} has also been shown to recover good estimates using a similar setup, using a control dataset based on the Current Population Survey (CPS-1), employing all pairwise interactions and squared terms for continuous variables, amounting to 52 covariates.

\begin{figure}[!h]
  \centering
  \caption{Estimating the Effect of a Job Training Program from Partially Observational Data}
  \label{fig:DW}
  \includegraphics[scale=.8]{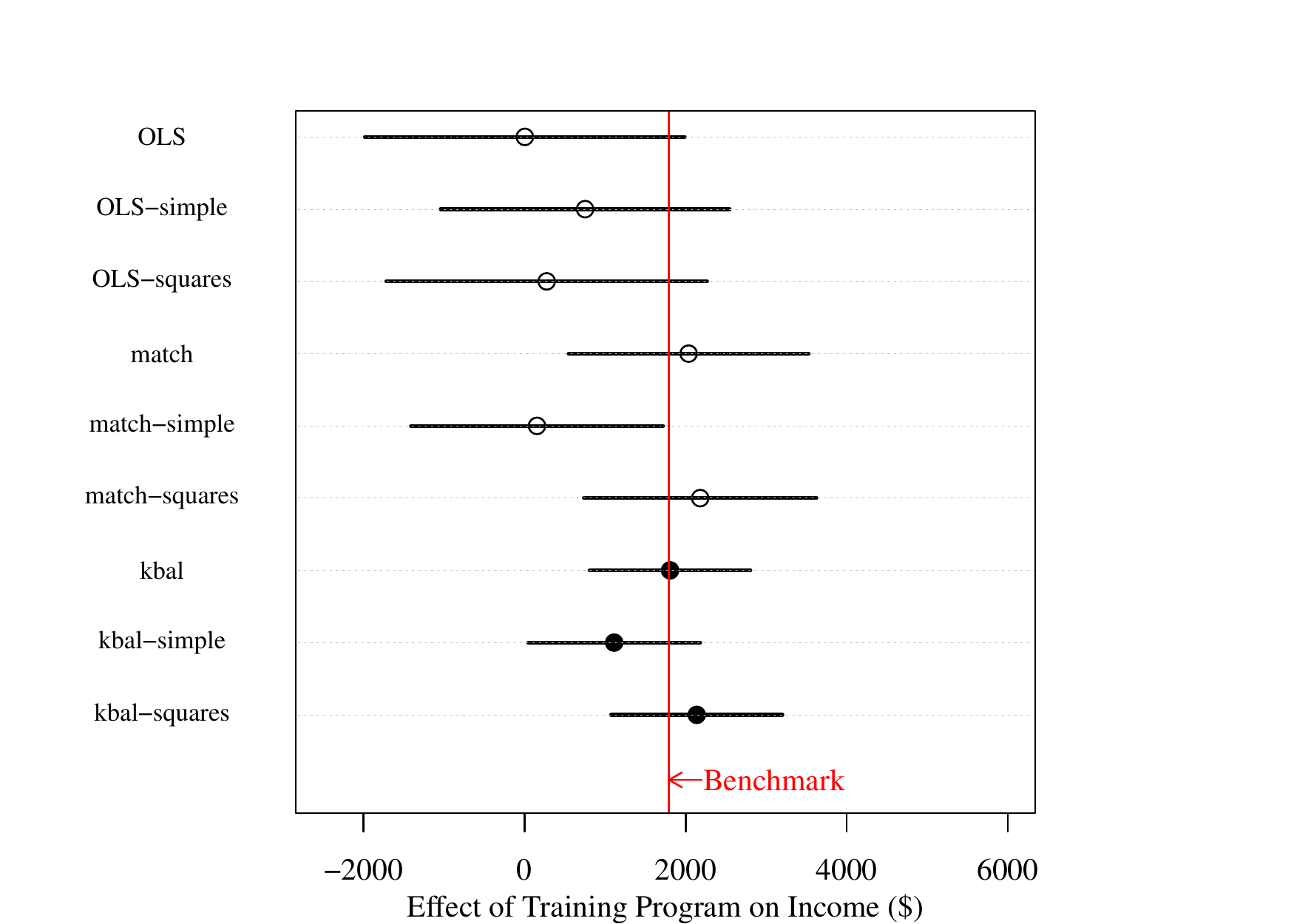}
\subcaption*{\footnotesize{Reanalysis of  \cite{dehejia1999causal}, estimating the effect of a job training program on income. Three procedures are used: linear regression (\emph{OLS}), Mahalanobis distance matching (\emph{Match}), and kernel balancing (\emph{kbal}). For each, three sets of covariates are attempted:  the standard set of 10 covariates described in the text, a reduced set (\emph{simple}) including only the seven of these that are not transforms of other variables, and an expanded set (\emph{squares}) including the 10 standard covariates plus squares of the three continuous variables. While \emph{OLS} and \emph{match} perform reasonably well, both are sensitive to specification. The best OLS estimate (\emph{OLS-simple}) still under-estimates the \$1794 benchmark by \$1042, while the best matching estimate (\emph{match-squares}) is off by \$387.
Kernel balancing performs reasonably well on all three specification, and the standard specification, $\emph{kbal}$, produces an estimate of \$1807, within \$13 of the benchmark.}}
\end{figure}

Figure \ref{fig:DW} reports results from a variety of estimation procedures and specifications. Three procedures are used: linear regression (\emph{OLS}), Mahalanobis distance matching (\emph{match}), and kernel balancing (\emph{kbal}). For \emph{match} and \emph{kbal}, estimate are produced by simple difference in means on the matched/reweighted sample. For comparability, all three approaches use simple standard errors that ignores any pre-processing stage, i.e. the	usual ``fixed'' weight standard errors.

For each method, three sets of covariates are attempted: the ``standard'' set of 10 covariates described above; a reduced set (\emph{simple}) including only the seven of these that are ``original'' variables, not transforms of others; and an expanded set (\emph{squares}) including the 10 standard covariates plus squares of the three continuous variables. 

Figure \ref{fig:DW} shows that the OLS estimates vary widely by specification, and even the estimate closest to the benchmark, \$1794, is incorrect by \$1042. Mahalahobis distance matching performs better, though remains somewhat specification dependent, with its best estimate (\emph{match-squares}) falling within \$387 of the benchmark. Finally, kernel balancing performs well over the three specification, with no estimate more than \$681 from the benchmark, and the standard specification, $\emph{kbal}$, producing an estimate of \$1807, within \$13 of the benchmark.

Perhaps the most important benefit of kernel balancing in this example is its relative insensitivity to  specification.  Matching performs well when the researcher knows to seek balance on particular non-linear functions -- specifically, on ``unemployment'' in 1974 and 1975, which are actually indicators for zero income. This transform of the original covariates is included in both \emph{match\_squares} and \emph{match} specifications. However in the \emph{match\_simple}, where only untransformed variables are used, matching produces a very different estimate, with a confidence interval that actually excludes the benchmark. By contrast, the kernel balancing estimates are closer to the benchmark in each case, less sensitive to specification, and simultaneously show less uncertainty. This reduced variance relative to other methods (where all methods use the ``fixed weights'' approach for comparability here) is likely due to the improved finite sample balance on characteristics influencing the outcome (see \cite{ho2007matching} for related discussion on improved efficiency due to pre-processing).

Further investigating the kernel balancing solution reveals additional details. We can see that balance is difficult to achieve in this example, in the sense that it requires focusing on a relatively small portion of the original control sample.  At the solution achieved by kernel balancing on the original variables alone ($kbal\_simple$), 90\% of the weight for controls is taken from just 98 units (reported automatically by \texttt{kbal}). So much weights falls to so few observations due to large differences between the treated and control samples. In examining why this is, the unemployment variable reveals its value: while 72\% of the treated are unemployed in either 1974 or 1975, only 12\% of controls are unemployed in either year. Using the $L_1$ measure described above, interpetable as either remaining imbalance on $\mathbf{K}$ or smoothed multivariate density imbalance, we get a value of $L_1=0.41$ prior to weighting. This indicates a considerable gap between the heights of the (smoothed) densities of the treated and control as evaluated at each datapoint. This is reduced to just $L_1=0.0016$ by kernel balancing. The choice of $r$ selected by the procedure was $45$ dimensions. With this choice, principal components accounting for $99.7$\% of the total variance of $\mathbf{K}$ are balanced upon. 

\section{Conclusions}\label{conclusions}
In the ongoing quest to reliably infer causal quantities from observational data, the primary challenge often remains ensuring that there are no unobserved confounders in a given identification scenario, so that assumptions such as Assumption \ref{weakSOO} are plausible.  However, even then, the mechanics of conditioning on observables to estimate causal effects remains non-trivial. Matching, covariate balancing weights, and propensity score weighting each seek to make the multivariate distribution of covariates for the untreated equal to that of the treated.  If any function of the observables that monotonically influences the non-treatment outcome persists in having a different mean for the treated and controls, the resulting estimates will be biased. Unfortunately, the investigator is not generally aware of all the functions of the covariates that may influence the outcome, making it difficult to guard against this possibility. 

However, unbiasedly estimating the SATT requires only that $\frac{1}{N_1} \sum_{i:D_i=1} Y_{0i} =\frac{1}{N_1} \sum_{i:D_i=0} Y_{0i}$, or ``mean balance on $Y_{0i}$''. Kernel balancing achieves this goal by working with the kernel matrix, $\mathbf{K}$, rather than the original covariates, $X$. It finds weights on the controls to make the weighted average row of $\mathbf{K}$ for the controls equal to the average row of $\mathbf{K}$ for the treated. Mean balance on these features implies mean balance on a large set of smooth functions of $X$. This includes all functions that can be formed by the superposition of Gaussians placed over each observation in the covariate space -- a very flexible space of functions that fits smooth functions particularly well in even smaller samples. The assumption that $\E[Y_{0i}|X_i]$ is among these functions is far more plausible than the assumption that it is linear in the original $X$, even if the investigator is careful enough to include higher-order terms among these $X$'s. Moreover as $N$ grows large, $\E[Y_{0i}|X_i]$ can be increasingly well accomodaed in this space.

While mean balance on $Y_{0i}$ is the principle goal, kernel balancing also implies that a particular kernel-based smoother for the multivariate densities is equal for the treated and control, as evaluated at every observation. Insofar as this is a reasonable density estimate, kernel balancing thus achieves what matching and covariate balancing estimators seek to achieve. These weights are also equivalent to a stabilized inverse propensity score weight that does not require an explicit model for the propensity score. This smoothed multivariate balance is achieved in a given sample, not just in expectation as is the case with traditional propensity score estimation. Thus, while focusing first on the minimum requirement for unbiased SATT estimation, the method also achieves the goals for which matching, weighting, and propensity score have traditionally been employed.  

Kernel balancing performs well in a reanalysis of \cite{dehejia1999causal}, a widely used benchmark for covariate adjustment in causal inference. At it's default values, with the covariates commonly used for this problem and no further specification choices, kernel balancing estimated an effect of \$1807 using the non-experimental control group, extremely close to the experimental 
benchmark of \$1794. Moreover, results are stable across specifications: getting an accurate result dow not depend upon foreknoweldge of non-linear functions that must be included to get a good result. Here agin, kernel balancing can be thought of as a principled choice of what functions of the covariates to acheive mean balance on, such that resulting ATT estimates are unbiased even when we do not know how exactly the covariates may influence the (non-treatment potential) outcome.

Numerous questions and challenges remain for future work.  First, $\mathbf{K}$ has dimensionality $N \times N$, which becomes unwieldy as $N$ grows large, posing a practical limit of tens of thousands of observations. Second, while the bootstrap is likely valid for obtaining confidence intervals that include uncertainty due to weight selection, further work on this is needed, and particularly on any approximations that may not be as computationally burdensome when $N$ is large. Finally, improvements may be possible on a number of implementation details, such as the choice of $b$, the optimization procedure for choosing the number of dimensions, alternate methods for dimension reduction on $\mathbf{K}$, and alternative methods for choosing the balancing weights that achieve mean balance on $\mathbf{K}$ while minimizing volatility. An implemnetation of this procedure using the choices described here is available in the \texttt{R} package \texttt{KBAL}, to be distributed on the CRAN repository upon acceptance of this paper. 

\clearpage

\singlespacing
\bibliography{kbalbib}

\clearpage
\section{Appendix}\label{appendix}

\subsection{Details of Motivating Example} \label{motivationdgp}
Details for the simulation are as follows: $War\,duration$ in years is distributed $max(1,N(7,9))$; $intensity$ in fatalities per year is distributed $Unif(100,10000)$. $fatalities$ is then computed as $intensity \cdot war\,duration$. The treatment, $peacekeeping$ is assigned by a Bernoulli draw with probability $logit^{-1}(\frac{intensity}{5000} - 2)$, and the outcome $peace\,years=\frac{intensity}{2500}+\epsilon$, $\epsilon \sim N(0,0.004)$.

\subsection{Further Implementation Details}\label{entropydetails}
\subsubsection*{Choice of Discrepancy Measure}
A method is needed to find the weight vector $w$ such that $\frac{1}{N_1}\mathbf{K}_t \mathbf{1}_{N_1} = \mathbf{K}_c w$, while constraining the weights to be non-negative and sum to one. It is also desirable to do this with minimal variation in the weights, by some measure, and in particular to avoid large weights. Two natural candidates for this are empirical likelihood \citep{owen1988empirical}, and entropy balancing \citep{hainmueller2012entropy}, both special cases of Cressie-Read divergence from a uniform distribution \citep{cressie1984multinomial}. Other approaches such as those that explicitly minimize the variation in weights for a given degree of imbalance (e.g. \citealp{zubizarreta2015stable}) may be valuable as well. In the \texttt{kbal}, I utilize entropy balancing, which seeks to satisfy these conditions while maximizing the Shannon entropy, $\sum_i w_i log(w_i)$, implied by the weights, which is also (proportional to) the Kullback divergence entropy between the distribution of weights and a uniform distribution. See \cite{hainmueller2012entropy} and references therein for further discussion.

\subsubsection{Optimization over $r$}

As described in the text, balance is achieved on the first $r$ principal components of $\mathbf{K}$, thereby achieving balance on the reconstructed approximation $\Kr$ closest to $\mathbf{K}$ in the Frobenius norm and operator 2-norm senses. How should $r$ be chosen? Since the goal is to achieve $\overline{k_t} = \sum_{i:D=0} w_i k_i$, a natural imabalnce measure to judge the success of a set of weights would be $a ||\overline{k_t} - \sum_{i:D=0} w_i k_i||$ for some norm $||\cdot||$ and constant $a$. Following \citep{CEMjasa}, I choose the $L_1$ measure here. As discussed below, this can serve as a measure of both imbalance on $\mathbf{K}$ and multivariate density imbalance insofar as densities are estimated by the corresponding kernel smooth (see \ref{L1equivalence}). 

The choice of $r$ is then made by beginning with $r=1$ and increasing it until a minimum in imbalance as measured by 
$L_1 = \frac{1}{2}\sum_{i=1}^{N} |\overline{k_t} - \sum_{i:D=0} w_i k_i|^1$. Alternative choices of norm (such as $L_2$ produce very similar results). Typically, imbalance improves as $r$ initially rises, and then deteriorates once $r$ is too high and numerical instability begins to creep in. An illustration of the relationship $r$, $L_1$ and the balance achieved on unknown functions of $X$ is given in the appendix (Figure \ref{fig:pctvarK}). In practice, for the $r$ chosen in this way, the number of principal components balanced  upon generally accounts for well over 99\% of the variance of $\mathbf{K}$. 

\subsubsection{Equivalence of $K$-imbalance and smoothed multivariate density imbalance}\label{L1equivalence}
Recall that the choice he optimization procedure chooses the number of projections of $\mathbf{K}$ that must be balanced while seeking to minimize overall imbalance on $\mathbf{K}$. Minimizing an imbalance measure of the form $a ||\overline{k_t} - \sum_{i:D=0} w_i k_i||$ for some norm $||\cdot||$ is natural given the goal of mean balance on $\mathbf{K}$. Such a norm also provides a measure of continuous multivariate imbalance. Setting $a$ to $\frac{1}{\sqrt{2\pi b}}$ to obtain $||\frac{1}{N_1 \sqrt{2\pi b}}\mathbf{K_t}^{\top} \mathbf{1}_{N_1} -\frac{1}{\sqrt{2\pi b}}\mathbf{K_c}^{\top}w ||$ we see this equals $||\hat{p}_{D=1}(\mathbf{X})-\hat{p}_{w,D=0}(\mathbf{X})||$, a norm on the difference between the smoothed density estimators for the treated and (weighted) controls, evaluated at each observation in the dataset.  Hence, norms of the form $||\overline{k_t} - \sum_{i:D=0} w_i k_i||$ are especially useful to minimize during optimization, as done in the selection of $r$ here, because they both minimize imbalance in $\mathbf{K}$ and a reasonable measure of ``multivariate imbalance'', i.e. a norm over the different in multivariate densities for the treated and control. 

When interpreted as a difference between estimated densities, the $L_1$ version of this norm described above is very much analogous to the $L_1$ metric used in Coarsened Exact Matching \citep{CEMjasa}, but without requiring coarsening in order to construct discrete bins in the covariates space. 


\subsection{Proof of Unbiasedness (Theorem \ref{dimwunbiased})}\label{proofunbiased}
Theorem \ref{dimwunbiased} states that the weighted difference in means estimator using kernel balancing weights is unbiased for the 
sample average treatment effect on the treated (SATT) and the (population) ATT.  

The SATT is similar to the ATT, but computes the average differences between the treatment and non-treatment potential outcome of the treated units actually sampled, rather than the expectation over the population distribution for the treated. The SATT is thus a more natural immediate target for an estimator. 
\begin{align}
SATT &= \frac{1}{N_1} \sum_{i:D_i=1} Y_{1i} - \frac{1}{N_0} \sum_{i:D_i=0} Y_{0i}
\end{align}

Recall that the $\widehat{DIM}_w$ is defined as $\frac{1}{N_1} Y_{1i} - \sum_{D=0} w_i Y_{0i}$. Recall also that under the assumption $\E[Y_{0i}|X_i]=\phi(X_i)^{\top}\theta$ (Assumption \ref{Y0linear}), $Y_{0i}=\phi(X_i)^{\top}\theta + \epsilon_i$ for $\E[\epsilon_i|X_i]=0$.

Hence the error of the $\widehat{DIM}_w$ estimate for the SATT is then
\begin{align}
\widehat{DIM}_w  - SATT &= \frac{1}{N_1} \sum_{i:D_i=1} Y_{0i} - \sum_{D_i=0} w_i Y_{0i} \\
&= \frac{1}{N_1} \sum_{i:D_i=1} \left(\phi(X_i)^{\top} \theta + \epsilon_i  \right) - \sum_{i:D_i=0} w_i \left(\phi(X_i)^{\top}\theta + \epsilon_i  \right) \\
&=\theta^{\top} \frac{1}{N_1} \sum_{i:D_i=1} \phi(X_i) + \frac{1}{N_1} \sum_{i:D_i=1} \epsilon_i  - \theta^{\top} \sum_{i:D_i=0} w_i \phi(X_i) - \sum_{i:D_i=0} w_i \epsilon_i  \\
&= \theta^{\top} \left(\frac{1}{N_1} \sum_{i:D_i=1} \phi(X_i)- \sum_{i:D_i=0} w_i \phi(X_i)  \right)  + \frac{1}{N_1} \sum_{i:D_i=1} \epsilon_i - \sum_{i:D_i=0} w_i \epsilon_i \\
&= 0 + \frac{1}{N_1} \sum_{i:D_i=1} \epsilon_i - \sum_{i:D_i=0} w_i \epsilon_i 
\end{align}

The bias is the expectation of this quantity, 
\begin{align}
bias &= \E\left[\widehat{DIM}_w  - SATT \right] \\
&= \E \left[ \frac{1}{N_1} \sum_{i:D_i=1} \epsilon_i - \sum_{i:D_i=0} w_i \epsilon_i \right] = 0
\end{align}

\subsubsection*{Remarks}\label{finitesamplemisspec}
Note that $\E[SATT]=ATT$, and so unbiasedness of $\widehat{DIM}_w$ for the SATT also implies unbiasedness for the $ATT$. 

The assumption that $\E[Y_{0i}|X_i]=\phi(X_i)^{\top}\theta$ is innocuous as $N \rightarrow \infty$, because the universal representation property of the Gaussian kernel ensures that the space of functions spanned by $\phi(X_i)^{\top}\theta$, which has representation $f(x_i)=\sum_j \alpha_j k(X_j,X_i)$, includes all continuous function. However, in finite samples the quality of approximation is limited. Imagine the superposition of Gaussians view of this functions space: with too few observations, there are limits to the shapes that can be built by placing Gaussians at each observation and rescaling them. Even though highly non-linear, non-additive functions can still be well modeled with relatively small samples (see \citealp{krlspaper}), we may still wish to know how finite samples behave in terms of potential bias.  Suppose that in truth, $\E[Y_{0i}|X_i]= \phi(X_i)^{\top}\theta + h(X_i) + \epsilon_i$, where $h(X_i)$ is the misspecification error, an additive component that cannot be captured by $\phi(X_i)^{\top}\theta$ using the sample available and by definition orthogonal to the span of $\phi(X_i)$. In this case, the difference between $\widehat{DIM}_w$ and the SATT becomes 
\begin{align} 
\widehat{DIM}_w  - SATT = \frac{1}{N_1} \sum_{i:D_i=1} \epsilon_i - \sum_{i:D_i=0} w_i \epsilon_i +
\frac{1}{N_1} \sum_{i:D_i=1} h(X_i) - \sum_{i:D_i=0} w_i h(X_i)  
\end{align}

Notice that bias due to misspecification occurs only if $h(X_i)$ has different means for the treated and controls (after weighting). That is, even if in a small sample $\E[Y_{0i}|X_i]$ cannot be well approximated, this is only problematic if the misspecification error, $h(X_i)$ is correlated with the treatment assignment after adjusting for differences on the other covariates through weighting. This is analogous to the biased caused by omitted variables in regression models.

\subsection{Balance in $\E[\phi(X_i)]$ implies balance in $\E[Y_{0i}]$}\label{populationbalance}
The main text focuses principally on SATT estimation, and the implications of obtaining balance on $\phi(X_i)$ in the finite sample. However working with populations instead, we note that obtaining $\E[\phi(X_i)|D_i=1]=\E_w[\phi(X_i)|D_i=0]$ also implies $\E[Y_{0i}|D_i=1]=\E_w[Y_{0i}|D_i=0]$, where $\E_w[\cdot]$ designates an expectation taken over the w-weighted distribution of $X$:
\begin{align}
\E[Y_{0i}|D=1] &= \E_x \left[\E[Y_{0i}|X,D=1]\right] \\
&= \theta^{\top} \int \phi(x) p(x|D=1)dx \\
&= \theta^{\top} \E[\phi(x)|D=1] \\
&& \nonumber \\
\E_w[Y_{0i}|D=0] &= \E_{w,x}\left[\E[Y_{0i}|X, D=0]\right] \\
&= \theta^{\top} \int \phi(x) w p(x|D=0)dx \\
&= \theta^{\top} \E_w[\phi(x)|D=0]
\end{align}

Hence when balance of $\phi(X_i)$ for the treated and controls holds in expectations, we will have $\E[Y_{0i}|D_i=1]=\E_w[Y_{0i}|D_i=0]$, allowing a (weighted) difference in means to unbiasedly estimate the ATT.

\subsection{Proof of proposition \ref{prop:PhiBalance}}
Proposition \ref{prop:PhiBalance} states: that for the mean row of $\mathbf{K}$ among the treated, $\overline{k_t}=\frac{1}{N_1}\mathbf{K_t} \mathbf{1}_{N_1}$ and the weighted mean row of $\mathbf{K}$ among the controls given by $\overline{k_c}(w)=\frac{\sum_{i} w_i k_i \mathds{1}_{\{D_i=0\}}}{N_0}$, if $\overline{k_t}=\overline{k_c}(w)$, then $\overline{\phi_t}=\overline{\phi_c}$ where $\overline{\phi_t}=\frac{1}{N_1}\sum_{D_i=1}\phi(x_i)$ and $\overline{\phi_c}=\sum_{D_i=0}\phi(x_i)$.

This can be shown as follows.
\begin{align}
\overline{k_T} &= \sum_{i:D_i=0}w_i k_i \label{balance} \\
\frac{1}{N_1}\left[\sum_{i:D_i=1} k(X_i,X_1), \ldots, \sum_{i:D_i=1} k(X_i,X_N)\right] &= \left[\sum_{i:D_i=0} w_i k(X_i,X_1), \ldots, \sum_{i:D_i=0} w_i k(X_i,X_N)\right]  \\
\frac{1}{N_1}\sum_{i:D_i=1} [\langle \phi(X_i),\phi(X_1) \rangle, \cdots, \langle \phi(X_i),\phi(X_N)\rangle ] &= \sum_{i:D_i=0} w_i [\langle \phi(X_i),\phi(X_1) \rangle, \cdots, \langle \phi(X_i),\phi(X_N)\rangle ] \\
\frac{1}{N_1}\sum_{i:D_i=1}\langle \phi(X_i),\phi(X_j)\rangle &= \sum_{i:D_i=0} w_i \langle \phi(X_i),\phi(X_j)\rangle \mbox{, } \forall j \\
\langle \frac{1}{N_1}\sum_{i:D_i=1}\phi(X_i),\phi(X_j)\rangle &= \langle \sum_{i:D_i=0} w_i \phi(X_i),\phi(X_j)\rangle \label{avesimilar} \\
\langle \overline{\phi_t},\phi(X_j)\rangle &= \langle \sum_{i:D_i=0} w_i \phi(X_i),\phi(X_j)\rangle \label{nearlythere} \\
\overline{\phi_t} &= \sum_{i:D_i=0} w_i \phi(X_i) \label{phibalancedone}
\end{align}

\subsubsection{Remarks}
An intuitive interpretation of equation \ref{avesimilar} is that each unit $j$ is as close to the average treated unit as it is to the (weighted) average control unit, where distance is measured in the feature space $\phi(X)$. For the Gaussian kernel, $\langle \phi(X_i),\phi(X_j)\rangle$ is naturally interpretable as a similarity measure in the \textit{input} space, since this quantity equals $k(X_j,X_i)=e^{-\frac{||X_j-X_i||^2}{b}}$. However, $\langle \phi(X_i),\phi(X_j)\rangle$ or $k(X_i,X_j)$ is more generally interpretable as similarity in the feature space as well. Note the squared Euclidean distance between two points $X_i$ and $X_j$ after mapping into $\phi(\cdot)$ is: $\|\phi(X_i)-\phi(X_j)\|^2 = \langle \phi(X_i)-\phi(X_j),\phi(X_i)-\phi(X_j) \rangle = \langle \phi(X_i),\phi(X_i) \rangle + \langle \phi(X_j),\phi(X_j) \rangle - 2 \langle \phi(X_i),\phi(X_j)\rangle$. In the case of the Gaussian kernel,  $\langle \phi(X_i), \phi(X_i) \rangle=1$, so this distance reduces to $2(1-\langle \phi(X_i),\phi(X_j) \rangle)$. In this sense,  $\langle \phi(X_i),\phi(X_j) \rangle$ is as reasonable measure of similarity of position in the feature space, as it runs opposite to distance in this space. 

Relatedly, a discriminant method of classifying observations as treated or control based on whether they are closer to the centroid of the treated or the centroid of the controls in $\phi(X)$ would be unable to classify any point. 

\subsection{Proof of proposition \ref{prop:densityequalization}}
Proposition \ref{prop:densityequalization} states that for a density estimator for the treated, $\hat{f}_{X|D=1}$, and for the (weighted) controls, $\hat{f}_{X|D=0,w}$, both constructed with kernel $k$ with scale $b$, the choice of weights that ensures mean balance in the kernel matrix $\mathbf{K}$ also ensures $\hat{f}_{X|D=1}=\hat{f}_{X|D=0,w}$ at every location in $\mathcal{X}$ at which an observation is located.

As detailed in the main text, the expression $\frac{1}{N_1 \sqrt{2\pi b}}K_t \mathbf{1}_{N_1}$ places a multivariate standard normal density over each \textit{treated} observation, sums these to construct a smooth density estimator at all points in $\mathcal{X}$, and evaluates the height of that joint density estimate at each of the points found in the dataset. Likewise, $\frac{1}{N_0 \sqrt{2\pi b}}K_c \mathbf{1}_{N_0}$ estimates the density of the control units and returns its evaluated height at every datapoint in the dataset.

To reweight the controls would be to say that some units originally observed should be made more or less likely. This is achieved by changing the numerator of each weight $\frac{1}{N_0 \sqrt{2\pi b}}$ to some non-negative value other than 1. Letting the weights sum to 1 (rather than $N_0$), the reweighted density of the controls would be evaluated at each point in the dataset according to $\frac{1}{\sqrt{2\pi b}} K_c w$, for vector of weights $w$. If weights are selected so that this equals the density of the treated:
\begin{align}
\frac{1}{N_1 \sqrt{2\pi b}}\mathbf{K_t} \mathbf{1}_{\{N_1\}} &= \frac{1}{\sqrt{2\pi b}}\mathbf{K_c} w \nonumber \\
\frac{1}{N_1}\mathbf{K_t} \mathbf{1}_{\{N_1\}} &= \mathbf{K_c} w \nonumber \\
\overline{k_t} &= \mathbf{K_c}w \nonumber \\
 \overline{k_t} &= \overline{k_c(w)}
\end{align}
\noindent where the final line is the definition of mean balance in $\mathbf{K}$. Thus, the weights that achieve mean balance in $\mathbf{K}$ are precisely the right weights to achieve equivalence of the measured multivariate densities for the treated and controls at all points in the dataset.

\subsection{Derivation of $\phi(X_i)$ for Gaussian Kernel}\label{phivector}
While the functions linear in $\phi(X_i)$ corresponding to a Gaussian kernel can more easily be understood as those that can be formed by superposing Gaussian kernels over the observations, one may also explicitly construct features $\phi(X_i)$ consistent with the requirement that $K(X_i, X_j) = \langle \phi(X_i). \phi(X_j) \rangle$ for the standard inner-product. One simple approach is, setting $b=.5$ for convenience, yields:

\begin{align}
k(X_i, X_j) &= exp(||X_i-X_j||^2) \\
&= exp(-X_i^2)exp(-X_j^2) exp(2X_i,X_j) \\
&= exp(-X_i^2)exp(-X_j^2) \sum_{d=0}^{\infty} \frac{2^d X_i^d X_j^d}{d!}
\end{align}

\noindent where the last line follows by a Taylor series expansion of $exp(2 X_i X_j)$.  Finally the division of terms can be completed, as:

\begin{align}
k(X_i, X_j) &= \sum_{d=0}^{\infty} \sqrt{\frac{2^d}{d!}}exp(-X_i^2 X_i^d) \sqrt{\frac{2^d}{d!}}exp(-X_i^2 X_i^d) 
\end{align}

This is simply an inner product of two infinite-dimensional vectors of the form
\begin{align}
\phi(X_i) &= \left[ \sqrt{\frac{2^0}{0!}}exp(-X_i^2 X_i^0),\;\;\sqrt{\frac{2^1}{1!}}exp(-X_i^2 X_i^1),\;\;...,\;\;  \sqrt{\frac{2^\infty}{\infty!}}exp(-X_i^2 X_i^\infty)  \right] 
\end{align}

\begin{figure}[!h]
  \centering
  \caption{First five values of $\phi(X)$ at varying values of $X$} 
  \label{fig:firstfive}
  \includegraphics[scale=.8]{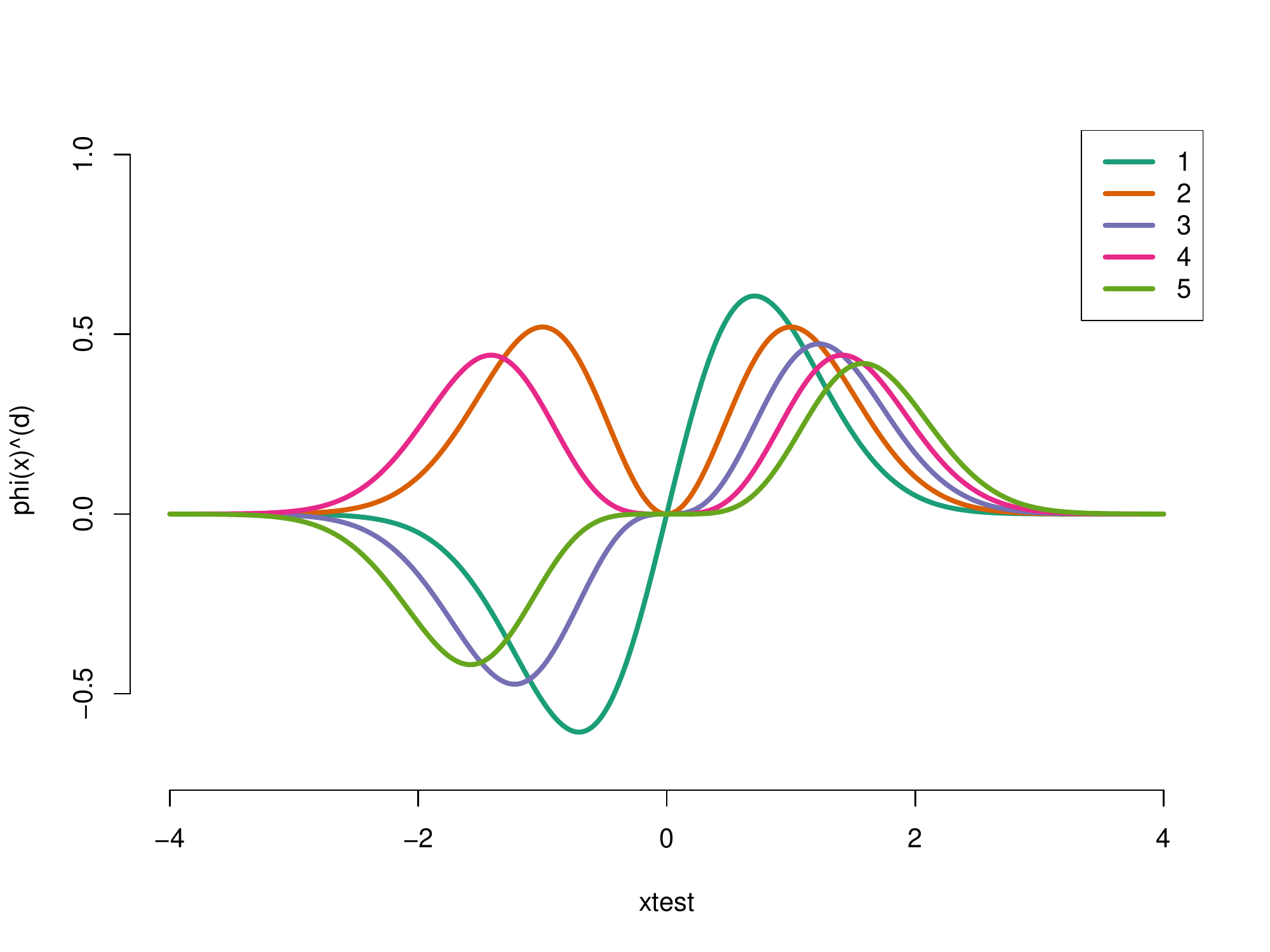}
\subcaption*{\footnotesize{Explicit view of $\phi(X_i)$} for one choice of $\phi(X_i)$ consistent with $K(X_i,X_j)=\langle \phi(X_i),\phi(X_j) \rangle$ for a Gaussian kernel $K$ as described in Equation \ref{phivector}.}
\end{figure}

Figure \ref{fig:firstfive} considers a one dimensional covariate, $X$, and shows what value each of the first 5 of these features would have at various values of $X$. 

\subsubsection{Density Equalization Illustration}

\begin{figure}[!hbt]
  \centering
  \caption{Density-Equalizing Property of Kernel Balancing}
  \label{fig:densitycomparison}
  \vspace{-.4in}
  \includegraphics[scale=.5]{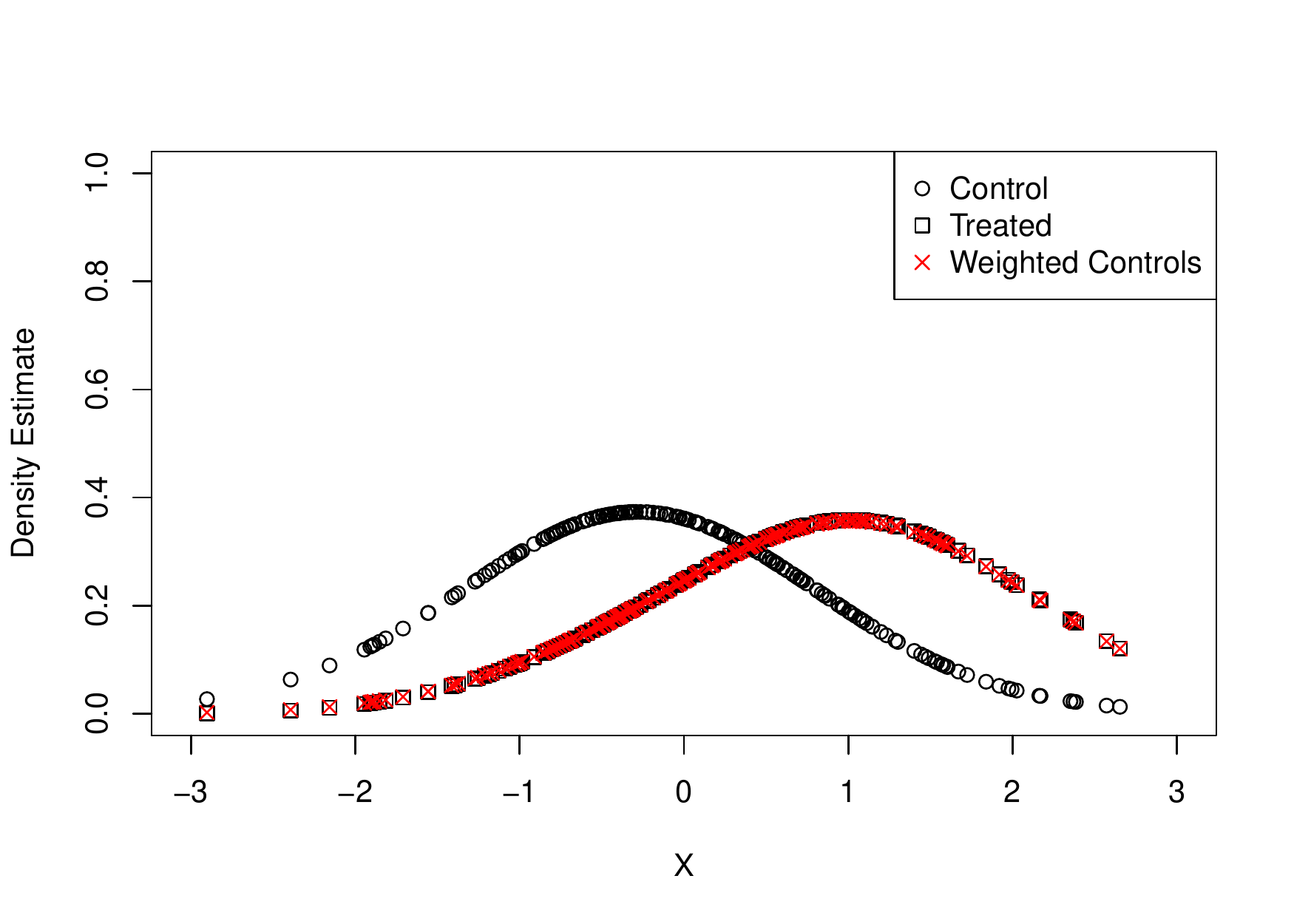}
\subcaption*{\scriptsize{Plot showing the density-equalization property of kernel balancing. For 200 observations of $X \sim N(0,1)$, treatment is assigned according to $Pr(treatment)=1/(1+exp(2-2X))$, producing approximately two control units for each treated unit. Black squares indicate the density of the treated, as evaluated at each observation's location in the dataset (and given the choice of kernel and $b$). Black circles indicate the density of (unweighted) controls. The treated and control are seen to be drawn from different distributions, owing to the treatment assignment process. Red x's show the new density of the controls, after weighting by \texttt{kbal}. The reweighted density is nearly indistinguishable from the density of the treated, owing to the density equalization property of kernel balancing.
}}
\end{figure}

This example visualized the density estimates produced internally by kernel balancing using linear combinations of $\mathbf{K}$ as described above. Suppose $X$ contains 200 observations from a standard normal distribution. Units are assigned to treatment with probability $1/(1+exp(2-2X))$, which produces approximately 2 control units for each treated unit.  Figure \ref{fig:densitycomparison} shows the resulting density plots, using density estimates provided by \texttt{kbal} in which the density of the treated is given by $\frac{1}{N_1 \sqrt{2\pi b}}\mathbf{K_t}\mathbf{1}_{N_1}$ and the density of the controls is given by $\frac{1}{N_0 \sqrt{2\pi b}}\mathbf{K_c}\mathbf{1}_{N_0}$.  As shown, the density estimates for the treated at each observations $X$ position (black squares) is initially very different from the density estimates for the controls taken at each observation (black circles). After weighting, however, the new density of the controls as measured at each observation (red \texttt{x}) matches that of the treated almost exactly.

Note that in multidimensional examples, the density becomes more difficult to visualize across each dimension, but it is still straightforward to compute and to think about the pointwise density estimates for the treated or control as measured at each observation's $X$ value. In contrast to binning approaches such as CEM,  equalizing density functions continuously in this way avoids difficult or arbitrary binning decisions, is tolerant of high dimensional data, and smoothly matches the densities in a continous fashion, resolving the within-bin discrepancies implied by CEM.

\subsubsection{$L_1$, imbalance, and $r$}
Recall that kernel balancing does not directly achieve mean balance on $\mathbf{K}$, but rather on the first $r$ factors of $\mathbf{K}$ as determines by principal components analysis. This example examines the efficacy of this approach in minimizing the $L_1$ loss, and in minimizing imbalance on an unknown function of the data. Suppose we have 500 observations and 5 covariates, each with a standard normal distribution.  Let $z=\sqrt{x_1^2+x_2^2}$. This function impacts treatment assignment, with the probability of treatment being given by $logit^{-1}(z-2)$, which produces approximately two control units for each treated unit.

\begin{figure}[!hbt]
  \centering
  \caption{$L_1$ distance and imbalance on an unknown confounder, by $r$}
  \label{fig:pctvarK}
  \vspace{-.4in}
  \includegraphics[scale=.8]{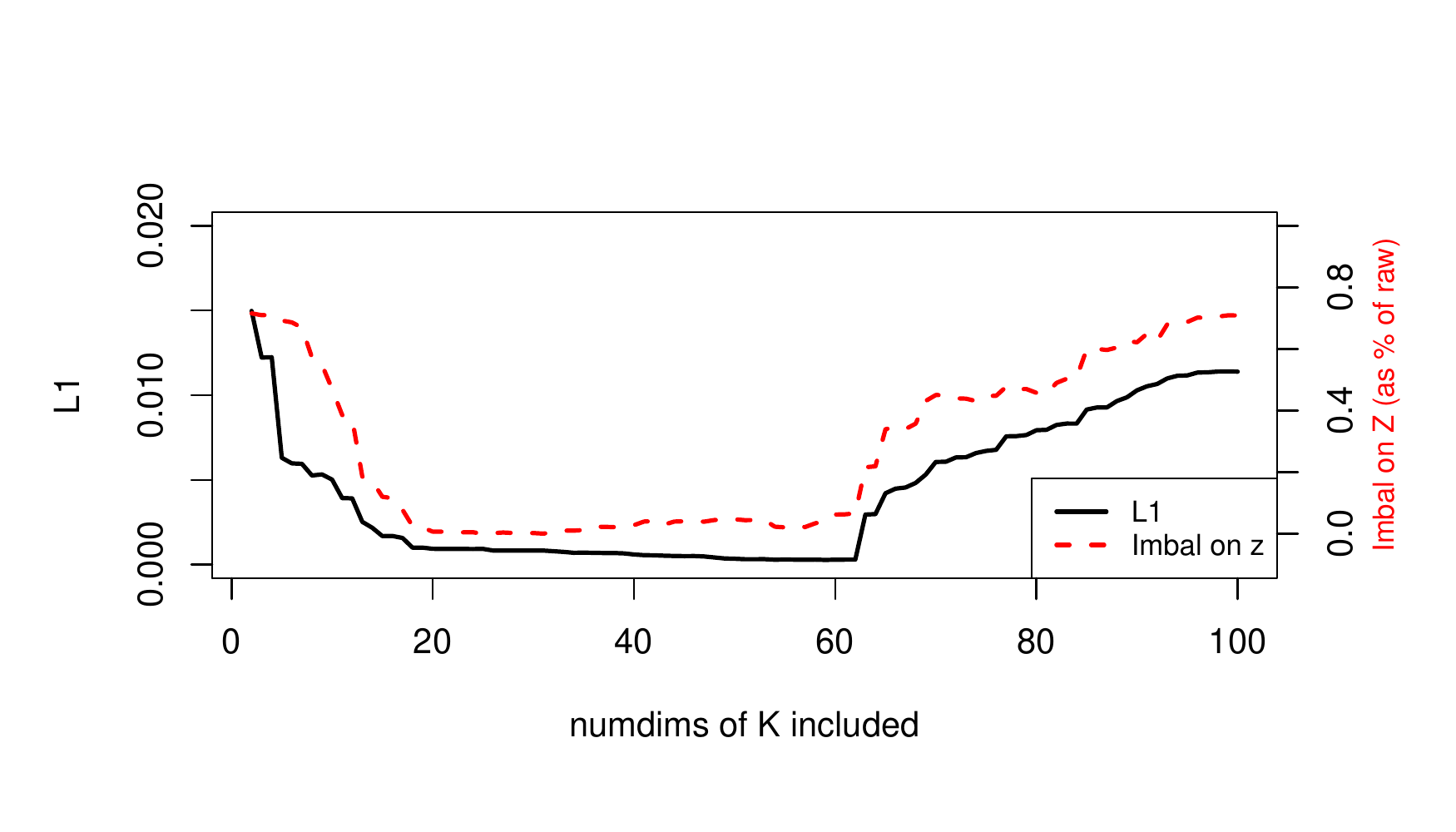}

\subcaption*{\scriptsize{This example shows the relationship between the number of components of $\mathbf{K}$ that get balanced upon ($r$), the multivariate imbalance ($L_1$), and balance on confounder $z$. $L_1$ generally improves as $r$ is increased at first, but beyond approximately 50 dimensions, numerical instability produces less desirable results and a higher $L_1$ imbalance. While the confounder represented by $z$ in this case would generally be unobservable, balance on $z$ is optimized where $L_1$ finds its minimum, which is observable.}}
\end{figure}

In Figure  \ref{fig:pctvarK}, the value of $r$ -- the number of factors of $\mathbf{K}$ retained for purposes of balancing  -- is increased from a minimum of 2 up to 100. As expected, both $L_1$ and the mean imbalance on $z$ taken after weighting improve as $r$ is first increased, and then worsen beyond some choice of $r$.  Most importantly, while the balance on $z$ is unobservable in the case of unknown confounders, $L_1$ is observable, and improvements in $L_1$ track very closely to improvements in the balance of $z$. Accordingly, selecting $r$ to minimize $L_1$ appears to be a viable strategy for selecting the value that also minimizes imbalance on unseen functions of the data.

\subsection{Inverse Propensity Score Weights as Multivariate Density Equalization}\label{appendix:ipw}
It is useful to show more explicitly the role played by inverse propensity score weights in estimating the ATT, as this leads to an appreciation of how these weights relate to multivariate density equalization, and the sense in which they are equivalent to the kernel balancing weights despite flowing from different initial goals.
 
Under Assumption \ref{weakSOO}, the ATT can be re-written:
\begin{align}
ATT &= \E[Y_{1i}|D_i=1]-\E[Y_{0i}|D_i=1] \\
&= \int \E[Y_{1i}|D_i=1,x]p(x|D_i=1)dx -\int \E[Y_{0i}|D_i=1,x]p(x|D_i=1)dx \\
&= \int \E[Y_{1i}|D_i=1,x]p(x|D_i=1)dx -\int \E[Y_{0i}|D_i=0,x]p(x|D_i=1)dx \label{almostID}
\end{align}

Expression \ref{almostID} is identifiable in the sense that we only require treatment potential outcomes from the treated units, and non-treatment potential outcomes from the non-treated units.  However, it remains problematic because it requires averaging outcomes from control units over the distribution of $X$ for the treated, $p(x|D_i=1)$, which is not the distribution of the control units in the sample. Specifically, the difference in means estimand, 
\begin{align}
\text{DIM} &= \E[Y_{1i}|D_i=1]-\E[Y_{0i}|D_i=0] \\
&= \int \E[Y_{1i}|D_i=1,x]p(x|D_i=1)dx -\int \E[Y_{0i}|D_i=1,x]p(x|D_i=0)dx \label{DIM2}
\end{align}

\noindent differs from the ATT in its second term, because it averages over the outcomes of non-treated units at their natural density in $X$, $p(x|D_i=0)$. To address this, consider a weighted difference in means estimand, 
\begin{align}
\text{DIM}_w &= \E[Y_{1i}|D_i=1]-\E_w[Y_{0i}|D_i=0] \\
&= \int \E[Y_{1i}|D_i=1,x]p(x|D_i=1)dx -\int w_i \E[Y_{0i}|D_i=1,x] p(x|D_i=0)dx 
\end{align} 

\noindent where $w_i$ is a function of $X$ that allows us to upweight or downweight control units. The difference between expression \ref{almostID} and \ref{DIM2} can be resolved by choosing weights
\begin{align}
w_i &= \frac{p(x|D_i=1)}{p(x|D_i=0)} \label{IPW2}
\end{align}

Through Bayes theorem, we can replace the class densities in this expression with more familiar propensity scores to obtain $w_i=\frac{p(D_i=1|x)p(D_i=0)}{p(D_i=0)|x)p(D_i=1)}$. For the control units ($D_i=0$), this is $w_i = \frac{p(D_i)}{p(D_i|X_i)}\frac{1-p(D_i|X_i)}{1-p(D_i)}$. These are the stabilized inverse propensity scores one would apply to the control units to estimate the ATT.  These weights, if properly estimated, ensure that the whole distribution of $X$ for the control units is adjusted to equal the distribution among the treated.

Note that in the form \ref{IPW}, it becomes clear that were we to adjust the sample to make treated and control groups have the same distribution of covariates, these weights would become constant and thus unnecessary. This is achieved, insofar as the smoothed multivariate densities on which kernel balancing obtains balance are reasonable approximations of the true densities.  In this sense, kernel balancing achieves the goals of inverse propensity score weighting, but has the advantage of avoiding any functional form assumption or direct estimation of the propensity score.

\subsection{Optional Trimming of the Treated} \label{Trimming}
In some cases, balance can be greatly improved with less variable (and thus more efficient) weights if the most difficult-to-match treated units are trimmed. In estimating an ATT, control units in areas with very low density of treated units can always be down-weighted (or dropped if the weight goes to zero), but treated units in areas unpopulated by control units pose a greater problem. These areas may prevent any suitable weighting solution, or may place extremely large (and thus ineffecient) weights on a small set of controls.

While estimates drawn from samples in which the treated are trimmed no longer represent the ATT with respect to the original population, they can be considered a local or sample average treatment effect within the remaining population. \cite{king2011comparative} refer similarly to a ``feasible sample average treatment effect on the treated'' (FSATT), based on only the treated units for which sufficiently close matches can be found. In any case, the discarded units can be characterized to learn how the inferential population has changed.

However, even when the investigator is willing to change the population of interest by trimming the treated, it is not always clear on what basis trimming should be done. In kernel balancing, trimming of the treated can be (optionally) employed by using the multivariate density interpretation given above. Specifically, the density estimators at all points is constructed using the kernel matrix. Then, treated units are trimmed if $\frac{p_{X|D=1}(x_i)}{p_{X|D=0}(x_i)}$ exceeds the parameter $trimratio$. The value of $trimratio$ can be set by the investigator based on qualitative considerations, inspection of the typical ratio of densities, a willingness to trim up to a certain percent of the sample, or performance on $L_1$. Whatever approach is taken to determine a suitable level of $trimratio$, \texttt{kbal} produces a list of the trimmed units, which the investigator can examine to determine how the inferential population has changed.

\subsection{Additional Example: Are Democracies Inferior Counterinsurgents?} \label{lyall}
Decades of research in international relations has argued that democracies are poor counterinsurgents (see \citealp{lyall2010democracies} for a review). Democracies, as the argument goes, are (1) sensitive to public backlash against wars that get more costly in blood or treasure than originally expected, (2) are unable to control the media in order to supress this backlash, and (3) often respect international prohibitions on brutal tactics that may be needed to obtain a quick victory. Each of these makes them more prone to withdrawal from countinsurgency operations, which often become long and bloody wars of attrition. Empirical work on this question was significantly advanced by \cite{lyall2010democracies}, who points out that previous work (1) often examined only democracies rather, than a universe of cases with variation on polity type, and (2) did little to overcome the non-random assignment of democracy, and particular, the selection effects by which democracies may choose to fight different types of counterinsurgencies than non-democracies.

\cite{lyall2010democracies} overcomes these shortcomings by constructing a dataset covering the period of 1800-2005, in which the polity type of the countinsurgent regimes vary.  Matching is then used to adjust for observable differences between the conflicts selected by democracies and non-democracies, using one-to-one nearest neighbor matching on a series of covariates. These covariates are: a dummy for whether the counterinsurgent is an occupier ($occupier$), a measure of support and sanctuary for insurgents from neighboring countries ($support$), a measure of state power ($power$), mechanization of the military ($mechanized$), $elevation$, $distance$ from the state capital to the war zone, a dummy for whether a state is in the first two years of independence ($new\,state$), a $cold\,war$ dummy, the number of $languages$ spoken in the country, and the $year$ in which the conflict began.

In a battery of analyses with varying modeling approaches, \cite{lyall2010democracies} finds that democracy, measured as a polity score of at least 7 in the specifications replicated here, has no relationship to success or failure in counter insurgency, either in the raw data or in the matched sample.

While the credibility of this estimate as a causal quantity depends on the absence of unobserved confounders, we can nevertheless assess whether the procedures used to adjust for observed covariates were sufficient, or whether an inability to achieve mean balance on some functions of the covariates may have led to bias even in the absence of unobserved confounders.

Here I reexamine these findings using the post-1945 portion of the data, which includes 35 counterinsurgencies by democracies and 100 by non-democracies, and is used in many of the analyses in \cite{lyall2010democracies}. The 1945 period is the only one with complete data on the covariates used for balancing here, but is also the period in which the logic of democratic vulnerability is expected to be most relevant. 

\begin{figure}[!h]
  \centering
  \caption{Balance: Democracies vs. Non-democracies and the Counterinsurgencies they Fight}
  \label{fig:Lyall_bal}
  \includegraphics[scale=.8]{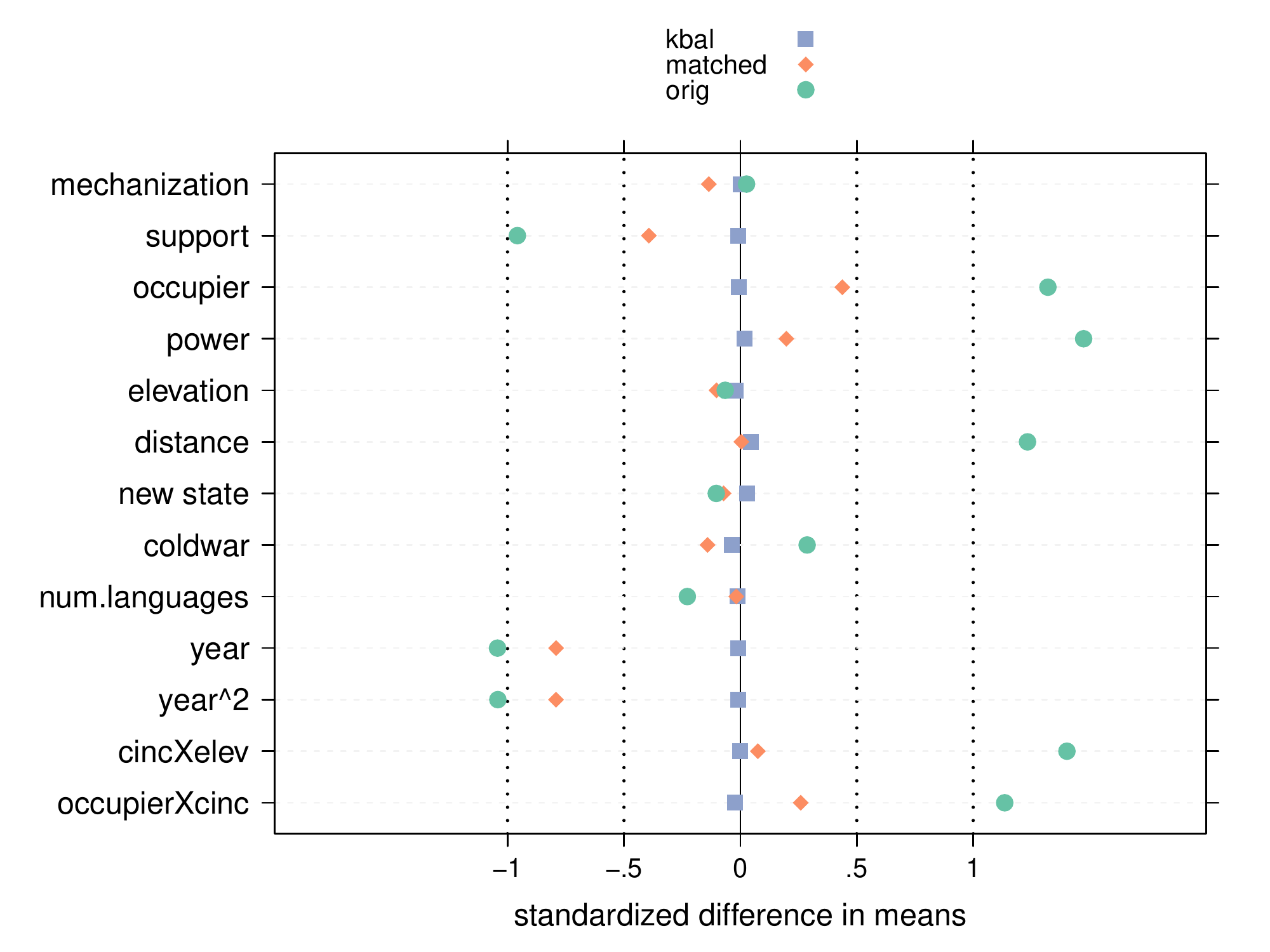}
\subcaption*{\footnotesize{Balance in post-1945 sample of \cite{lyall2010democracies}. Imbalance, measured as the difference in means divided by the standard deviation, is shown on the $x$- axis. Democracies (treated) and non-democracies (controls) vary widely on numerous covariates. The matched sample (diamonds) shows somewhat improved balance over the original sample, but imbalances remain on numerous characteristics. Balance is considerably improved by kernel balancing (squares). The rows at or above $year$ show imbalance on characteristics explicitly included in the balancing procedures. Those below $year$ show imbalance on characteristics not explicitly included.}}
\end{figure}

First, I assess balance. As shown in Figure \ref{fig:Lyall_bal}, numerous covariates are badly imbalanced in the original dataset (circles), where imbalance is measured on the $x$-axis by the standardized difference in means. This balance improves somewhat under matching (diamonds), but improves far more under kernel balancing (squares). Note that imbalance is shown both on the variables used in the matching/weighting algorithms (the first ten covariates up to and including $year$), as well as several others that were not explicitly included in the balancing procedure: $year^2$, and two multiplicative interactions that were particularly predicted of treatment status in the original data. Kernel balancing produces good balance on both the included covariates, and functions of them.

Next, I use the matched and weighted data to estimate the effect of democracy on counterinsurgency success. For this, I simply use linear probability models (LPM) to regress a dummy for victory (1) or defeat (0) on covariates according to five different specifications. While \cite{lyall2010democracies} used a number of other approaches, including logistic regression, some of these models suffer ``separation'' under the specifications attempted here. This causes observations and variables to effectively drop out of the analysis, producing variability in effect estimates that are due only to this artefact of logistic regression and not due to any meaningful change in the relationship among the variables. Linear models do not suffer this problem, and provide a well defined approximation to the conditional expectation function, allowing valid estimation of the changing probability of victory associated with changes in the treatement variable, $democracy$. The first three specifications used are (1) \emph{raw} regresses the outcome directly on $democracy$ without covariates (and is equivalent to difference-in-means);(2) \emph{orig} uses the same covariates as \cite{lyall2010democracies}, which are all those variables balanced on except for $year$, (3) \emph{time} reincludes $year$ as well as $year^2$ to flexibly model the effects of time. The final two models, \emph{occupier1} (4) and \emph{occupier2} (5), add flexibility by including interactions of $occupier$ with other variables in the model. These interactions were chosen because analysis with KRLS revealed that interactions with $occupier$ were particularly predictive of the outcome.

\begin{figure}[!h]
  \centering
  \caption{Effect of Democracy on Counterinsurgency Success}
    \label{fig:Lyall_ATT}
  \includegraphics[scale=.8]{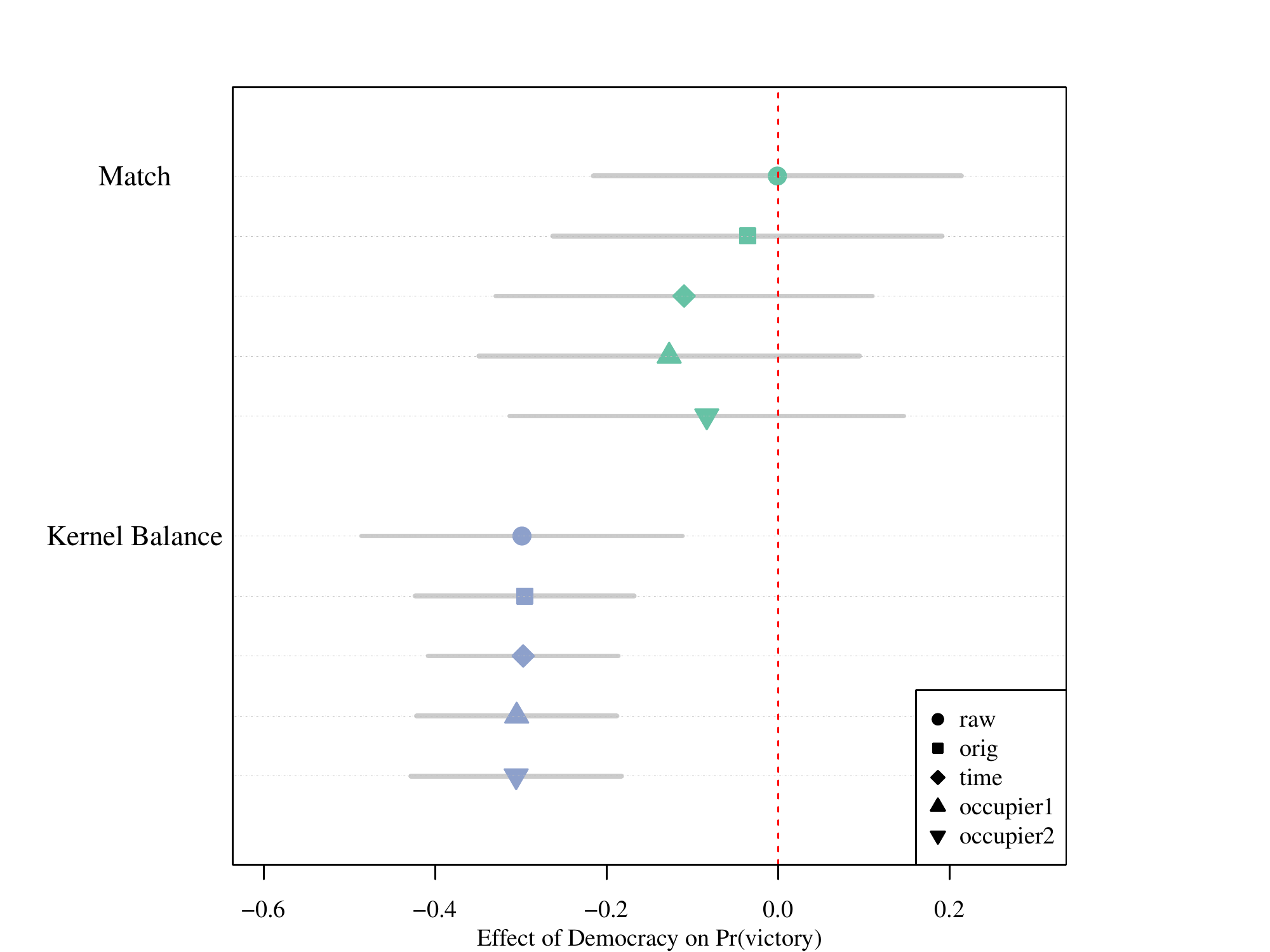}
\subcaption*{\footnotesize{Effect of democracy on counterinsurgency success in post-1945 sample of \cite{lyall2010democracies} using matching or kernel balancing for pre-processing followed by five different estimation procedures. Under matching, effect estimates remain highly variable, but none are significantly difference from zero.  Kernel balancing shows remarkably stable estimates over the five estimation procedures, even when no covariates are included (\emph{raw}). Results from kernel balancing are consistently in the -0.26 to -0.27 range and significantly different from zero, indicating that democracy is associated with a substantively large deficit in the ability to win counterinsurgencies.}}
\end{figure}

Figure \ref{fig:Lyall_ATT} shows results for the matched and kernel balanced samples with 95\% confidence intervals. Under matching, the effect varies considerably depending on the choice of model. No estimate is significantly different from zero, however. In stark contrast, kernel balancing producing estimates that are essentially invariant to the choice of model. Each kernel balancing estimate is between $-0.26$ and $-0.27$, indicating that democracy is associated with a 26 to 27 percentage point lower probability of success in fighting counterinsurgencies.  This is a very large effect, both statistically and substantively, given that the overall success rate is only 33\% in the post-1945 sample.

\subsection{Are democracies more selective?}
One puzzle regarding the claim that democracies are inferior counterinsurgents has been why democracies, whatever their weaknesses as counerinsurgents, are not also better able to ``select into'' conflicts they are more likely to win. The same qualities that are theorized to make democracies more susceptible to defeat against insurgents -- public accountability and media freedoms -- might also push democracies to more carefully select what counterinsurgency operations they  engage in.

The findings suggest that such a selection may occur. Specifically, the naive effect estimate obtained by a simple difference in mean probability of victory (on the unweighted sample) is -0.10 ($p=0.13$). Recall that this difference in means can be decomposed, 
 \begin{align}
 \E[Y(1)|D=1]-\E[Y(0)|D=0] &= \E[Y(1)|D=1]- \E[Y(0)|D=1]+\E[Y(0)|D=1]-\E[Y(0)|D=0]\nonumber \\
                                           &= ATT + [\E[Y(0)|D=1]-\E[Y(0)|D=0]] \nonumber
 \end{align}

That is, the naive difference in means is the average treatment effect on the treated (had they fought in the same types of cases), plus a selection effect indicating how democracies and non-democracies differ in their probabilities of victory based only on fighting different types of cases (i.e. in the absence of any effect of democracy). Since we know the ATT estimate and the raw difference in means, we can estimate the selection effect to be about 17 percentage points more likely to end in victory. While simple, this decomposition suggests that democracies do choose counterinsurgencies somewhat ``wisely'', but are also less likely to win a given a counterinsurgency once this selection is accounted for.

\end{document}